\documentclass[11pt,english,onecolumn,draftcls]{IEEEtran}
\usepackage[T1]{fontenc}
\usepackage[latin9]{inputenc}
\usepackage{units}
\usepackage{amsthm}
\usepackage{amsmath}
\usepackage{amssymb}
\usepackage{mathdots}
\usepackage{graphicx}
\usepackage{setspace}
\PassOptionsToPackage{version=3}{mhchem}
\usepackage{mhchem}
\usepackage{esint}
\makeatletter
\theoremstyle{plain}
\newtheorem{thm}{\protect\theoremname}
\theoremstyle{plain}
\newtheorem{prop}[thm]{\protect\propositionname}
\theoremstyle{remark}
\newtheorem{rem}[thm]{\protect\remarkname}
\theoremstyle{plain}
\newtheorem{lem}[thm]{\protect\lemmaname}

\usepackage{hyperref}
\usepackage{breqn}
  
\DeclareMathOperator*{\argmax}{arg\,max}

\allowdisplaybreaks

\global\long\def\s[#1]{\textnormal{\scriptsize #1}}
\global\long\def\st[#1]{\mathrm{\scriptsize #1}}

\global\long\def\P{\mathbb{P}}
\global\long\def\E{\mathbb{E}}
\global\long\def\I{\mathbb{I}}

\global\long\def\d{\mathrm{d}}

\global\long\def\teq{\triangleq}
\global\long\def\dfn{\teq}

\global\long\def\trre[#1,#2]{\overset{{\scriptstyle (#2)}}{#1}} % transition explained with reason
\global\long\def\bin[#1,#2]{\mathbb{B}[#1;#2]} %Binary representation of an element in a set
\global\long\def\dec[#1]{\mathbb{D}_{#1}} %Decimal representation of bits

\author{
\authorblockN{Nir Weinberger and Neri Merhav}

\authorblockA{Dept. of Electrical Engineering\\
  	    Technion - Israel Institute of Technology\\
Technion City, Haifa 3200004, Israel
} \\
\authorblockA{\{nirwein@campus, merhav@ee\}.technion.ac.il}\\
}
\makeatother

\usepackage{babel}
\providecommand{\lemmaname}{Lemma}
\providecommand{\propositionname}{Proposition}
\providecommand{\remarkname}{Remark}
\providecommand{\theoremname}{Theorem}

\begin{document}

\title{Lower Bounds on Parameter Modulation\textendash Estimation Under
Bandwidth Constraints}

\maketitle
\renewcommand\[{\begin{equation}}
\renewcommand\]{\end{equation}}
\thispagestyle{empty}
\begin{abstract}
We consider the problem of modulating the value of a parameter onto
a band-limited signal to be transmitted over a continuous-time, additive
white Gaussian noise (AWGN) channel, and estimating this parameter
at the receiver. The performance is measured by the mean power-$\alpha$
error (MP$\alpha$E), which is defined as the worst-case $\alpha$-th
order moment of the absolute estimation error. The optimal exponential
decay rate of the MP$\alpha$E as a function of the transmission time,
is investigated. Two upper (converse) bounds on the MP$\alpha$E exponent
are derived, on the basis of known bounds for the AWGN channel of
inputs with unlimited bandwidth. The bounds are computed for typical
values of the error moment and the signal-to-noise ratio (SNR), and
the SNR asymptotics of the different bounds are analyzed. The new
bounds are compared to known converse and achievability bounds, which
were derived from channel coding considerations.\end{abstract}

\section{Introduction\label{sec:Introduction}}

The problem of \emph{waveform communication}, as termed in the classic
book by Wozencraft and Jacobs \cite[Chapter 8]{jacobs1965principles},
is about conveying the value of a continuous valued parameter to a
distant location, via a communication channel. Formally, at the input
of the channel, a modulator maps a parameter%
\footnote{The range of values the parameter may take is assumed $[0,1)$ for
reasons of convenience only, with no essential loss of generality.%
} $u\in[0,1)$ to a signal $\{s(t,u),\;0\leq t\leq T\}$, which is
transmitted over the continuous-time AWGN channel, under a power constraint,
$\frac{1}{T}\int_{0}^{T}s^{2}(t,u)\d t\leq P$. At the output of the
channel, an estimator processes the received signal, $y(t)=s(t,u)+z(t),\;0\leq t\leq T$,
to obtain an estimate of the parameter. Here, $z(t)$ is a Gaussian
white noise process with two-sided spectral density $\nicefrac{N_{0}}{2}$.

\begin{figure}
\begin{centering}
\includegraphics[scale=1.2]{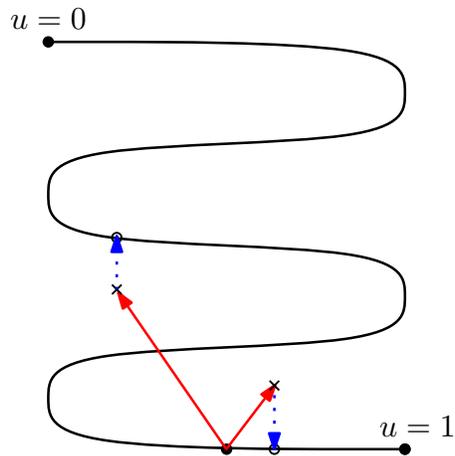}
\par\end{centering}

\protect\caption{The signal space with a locus obtained from a Shannon-Kotel'inkov
mapping \cite[Fig. 4]{shannon1949communication}. As the parameter
exhausts the interval $[0,1)$, the point on the signal space travels
from the top left point of the locus to its bottom right point. Dots
corresponds to actual parameter values. The solid (red) arrows represent
the effect of the noise, and the circles represent estimated parameter
values. The dotted (blue) arrows represent the estimation stage. \label{fig:Shannon-Kotel'inkov-mapping}}
\end{figure}

Such a modulation-estimation system can be depicted in a geometrical
way, as shown in Fig. \ref{fig:Shannon-Kotel'inkov-mapping}. As noticed
by Kotel'inkov \cite{Shannon_Kotelnikov} and Shannon \cite{shannon1949communication},
 the various signals $\{s(t,u),\;0\leq u<1\}$ can be represented
as vectors in some signal space. The modulator can therefore be viewed
as mapping the parameter into a point in the signal space, and as
the parameter exhausts its domain, a locus (possibly, discontinuous)
in the signal space is obtained. The additive noise then shifts the
transmitted point to a different point in the signal space, and the
estimator maps it back to a point on the locus, which in turn corresponds
to an estimated value of the parameter. The estimation performance
is usually evaluated by the $\alpha$-th order moment of the absolute
estimation error, which we term the \emph{mean power-$\alpha$ error}
(MP$\alpha$E). Most commonly, the mean square error (MSE) is used
($\alpha=2$). As can be discerned from Fig. \ref{fig:Shannon-Kotel'inkov-mapping},
the estimation error can be roughly categorized into two types: (i)\emph{
weak} \emph{noise errors}, which\emph{ }result in small estimation
errors, and are associated with the local, linearized behavior of
the locus (right red arrow in Fig. \ref{fig:Shannon-Kotel'inkov-mapping}),
and (ii)\emph{ anomalous errors}, that yield relatively large estimation
errors, which are associated with the twisted curvature of the locus
in the signal space for non-linear modulation systems (left red arrow
in Fig. \ref{fig:Shannon-Kotel'inkov-mapping}). A good communication
system should properly balance between the two types of errors. Nonetheless,
when the SNR falls below a certain threshold, the anomalous error
quickly dominates and the MP$\alpha$E becomes catastrophic. This
phenomenon is known as the \emph{threshold effect}, see, e.g., \cite{jacobs1965principles,Neri_threshold_effects},
and many references therein.%
\footnote{We refer the reader to \cite[Section 2]{Neri_threshold_effects} and
\cite[Section 2]{Shannon_Kotelnikov}, for a more detailed discussion
on the waveform communication problem.%
}

A natural question, for such systems, is how small can the MP$\alpha$E
be made for an arbitrary modulation-estimation system, operating over
transmission time $T$? As usual, answering this question exactly
is prohibitively complex, even in very low dimensions \cite{Shannon_Kotelnikov,chung2000construction}.
However, it turns out that if the modulator and estimator are designed
carefully, the MP$\alpha$E may decay exponentially with $T$, to
wit
\[
\sup_{u\in[0,1)}\E_{u}\left\{ \left|\hat{U}-u\right|^{\alpha}\right\} \approx e^{-E\cdot T}
\]
for some constant $E>0$, where $\hat{U}$ is the estimator%
\footnote{A more precise definition will be given in the sequel.%
} and $\E_{u}\{\cdot\}$ is the expectation operator with respect to
(w.r.t.) the channel noise, when the underlying parameter is $u$.
As we next review, the optimal exponential decay rate of the MP$\alpha$E
was investigated, in the same spirit that the optimal exponential
decay of the error probability was studied for the problem of channel
coding (e.g. \cite[Chapter 5]{gallager1968information} and \cite[Chapter 10]{csiszar2011information}). 

Most of the previous research has focused on the AWGN channel without
bandwidth constraints on the input signals. The goal of this paper
is to develop bounds on the MP$\alpha$E for band-limited input signals,
with emphasis on lower bounds. Since a lower bound on the MP$\alpha$E
is associated with an upper bound on its exponent and vice-versa,
then to avoid confusion, throughout the paper the term `converse bound'
will be used in the sense of an upper bound on the MP$\alpha$E exponent.
Similarly, the term `achievability bound' will be used for a lower
bound on the MP$\alpha$E exponent. Nevertheless, the terms `converse'
and `achievability' are only used here in a loose way, in the sense
that it does not necessarily imply that the lower bound on the exponent
coincides with the upper bound.

We begin with a short review on existing bounds for the continuous-time,
unlimited-bandwidth case. For achievability results, a few simple
systems were considered. In \cite[Chapter 8]{jacobs1965principles},
a frequency position modulation (FPM) system with a central frequency
and bandwidth that both increase exponentially with $T$, i.e., as
$\exp(RT)$, for some optimized $R$, was shown to achieve an exponential
decrease of the MSE according to $\exp(-\frac{P}{3N_{0}}T)$. In the
same spirit, a pulse position modulation (PPM) can be used, again,
with exponentially increasing bandwidth, to achieve the same exponent.
More recently, a modulation scheme which employs uniform quantization
of the parameter to $\exp(RT)$ values (where $R>0$ is again a design
parameter), followed by an optimal rate-$R$ channel code for AWGN
channel (i.e., its \emph{reliability function}), was shown to achieve
the same exponent (see \cite[Introduction]{Neri_modulation_estimation}).
A similar system will be discussed in Section \ref{sec:Converse-Bounds}. 

To assess the performance of the above schemes, converse bounds have
also been derived. On the face of it, as this problem lies in the
intersection between information theory and estimation theory, methods
from both fields are expected to have the potential to provide answers.
While estimation theory offers an ample of Bayesian and non-Bayesian
bounds \cite{van2013detection} (see also \cite[Introduction]{Ben-Haim}
and references therein for an overview), the vast majority of them
strongly depend on the specific modulator, and so, they are less useful
for us in the quest for \emph{universal} bounds, i.e., when there
is freedom to optimize the modulator. From the information-theoretic
perspective, one can view the parameter as an information source,
and assume that it is a random variable $U$, say, distributed uniformly
over $[0,1)$. The estimate $\hat{U}$, is then chosen to minimize
the average distortion, under a distortion measure defined as the
$\alpha$-th order moment of the absolute error. The MP$\alpha$E
is then the average distortion $D$ of this joint source-channel\emph{
}coding system, and, in principle, the data processing theorem (DPT)
\cite[Section 7.13]{Cover:2006:EIT:1146355} can be harnessed to obtain
a converse bound of the form $D\geq R^{-1}(C)$, where $R(D)$ is
the rate-distortion function of the source and $C$ is the channel
capacity. However, this bound may be too optimistic, since to achieve
this bound using a separation-based system, the source should be compressed
at a rate close to its rate-distortion function, which is impossible
when there is merely a single source symbol (scalar quantization).%
\footnote{The same is true for any given \emph{finite} dimension, that does
not grow with $T$.%
}

In the unlimited-bandwidth case, $C=\frac{P}{N_{0}}$, and while the
rate-distortion function is not known to have a closed form formula,
it can be lower bounded using Shannon's lower bound (e.g. \cite[Corollary 7.7.5]{viterbi2009principles},
\cite[Section 4.3.3]{berger1971rate})\textbf{ }as 
\[
R(D)\geq h(U)-\frac{1}{2}\log(2\pi eD)=-\frac{1}{2}\log(2\pi eD)
\]
where $h(U)=0$ is the differential entropy of $U$. Therefore, the
DPT lends itself to obtain a lower bound on the MSE, given by 
\[
D=\E(U-\hat{U})^{2}\geq\frac{1}{2\pi e}\exp(-2CT)=\frac{1}{2\pi e}\exp\left(-\frac{2P}{N_{0}}T\right).
\]

In \cite[Section 6]{zakai1975generalization} the idea of using a
DPT with generalized information measures \cite{Ziv_zakai_functionals},
which pertain to a general \emph{univariate} convex function, was
extended to \emph{multivariate} convex functions, and harvested in
order to obtain the improved bound of the exponential order of $\exp(-\frac{P}{N_{0}}T)$. 

In a different line of work, a more direct approach was taken, and
a lower bound on the MP$\alpha$E was developed from an analysis of
the channel coding system introduced above, namely, a modulation system
which maps a quantized value of the parameter to a codeword from a
channel code (or a signal from a \emph{signal set). }Rather complicated
arguments were used to obtain a converse bound which is valid for
any signal set. Research in this direction was initiated by Cohn in
his Ph.D. thesis \cite{Cohn}, who derived a lower bound of the exponential
order $\exp(-\frac{P}{2.89\cdot N_{0}}T)$ for the MSE ($\alpha=2$).
Later on, Burnashev \cite{Burnashev84,Burnashev85} has revised and
generalized Cohn's arguments, and his efforts eventually culminated
in \cite[Theorem 3]{Burnashev85}, which provides, among other results,
the lower exponential bound of the exponential order of $\exp(-\frac{P}{3N_{0}}T)$
for $\alpha=2$. As this converse bound coincides exponentially with
the achievability bound, then the optimal exponent is precisely characterized
for the unlimited-bandwidth AWGN channel. 

The exploration of universal bounds to modulation-estimation problem
was not confined only to AWGN channels and the MP$\alpha$E. In \cite[Section IV]{Some_lower_bounds},
a large deviations performance metric was considered, namely, the
exponential behavior of the probability that the estimation error
would exceed some threshold. This exponent was fully characterized
in \cite{Neri_modulation_estimation}: For an optimal communication
system, the probability that the absolute estimation error would exceed
$\exp(-RT)$ behaves exponentially as $\exp[-T\cdot E(R)]$, where
$E(R)$ is the reliability function of the channel.%
\footnote{The result in \cite{Neri_modulation_estimation} assumes an unlimited-bandwidth
AWGN channel, for which the reliability function is known exactly
(c.f. Remark \ref{rem: channel coding achievable for unlimited bandwidth channels}).
However, the proofs in \cite{Neri_modulation_estimation} are general,
and in fact pertain to any channel for which a reliability function
exists. %
}

The exponential behavior of the MP$\alpha$E discussed above for continuous-time
channels, holds when there is no limitation on the bandwidth of the
input signals. In \cite{Neri_estimation_DMC}, a converse bound and
an achievability bound on exponent of the MP$\alpha$E were derived,
for a discrete memoryless channel (in discrete-time), rather than
the AWGN channel (in continuous-time). In this paper, we consider
the problem of characterizing the maximal achievable exponent of the
MP$\alpha$E for the AWGN channel fed by a \emph{band-limited} input,
with emphasis on converse bounds. We are not aware of earlier works
that focus concretely on this setting.

As a simple benchmark, the DPT bound mentioned above can be adapted
to input signals band-limited to $W$, by simply replacing the capacity
of the unlimited-bandwidth case with the capacity of AWGN channel
with band-limited inputs, i.e., 
\begin{equation}
C=W\log\left(1+\frac{P}{N_{0}W}\right).\label{eq: capacity bandlimited channel}
\end{equation}
The resulting lower bound on the MP$\alpha$E has exponential order
of%
\footnote{The DPT bounds as stated in \eqref{eq: DPT band-limited} is suitable
for $\alpha=2$, since Shannon's lower bound was used for the MSE
distortion measure. To generalize it to other values of $\alpha$,
we recall that for difference distortion measures, Shannon's lower
bound is given by the entropy of the source minus the maximum entropy
\cite[Chapter 12]{Cover:2006:EIT:1146355} over all random variables
satisfying the distortion constraint. For a distortion measure of
the form $d(u,\hat{u})=|u-\hat{u}|^{\alpha}$ the maximum entropy
is obtained by a generalized Gaussian density with parameter $\alpha$,
i.e., $f(x)\sim\exp\{-\left|\frac{x}{s}\right|^{\alpha}\}$ where
$s$ is a scaling parameter. So, the Shannon lower bound in this case
is given by $h(U)+d_{\alpha}-\frac{1}{\alpha}\log D$, where $d_{\alpha}$
depends only on $\alpha$, and does not affect the exponential behavior
of the bound. This and \eqref{eq: capacity bandlimited channel} immediately
imply \eqref{eq: DPT band-limited}.%
} 
\begin{equation}
\exp\left[-T\cdot\alpha W\log\left(1+\frac{P}{N_{0}W}\right)\right].\label{eq: DPT band-limited}
\end{equation}
Thus, unlike the unlimited-bandwidth case, for which the MP$\alpha$E
scales linearly with $\frac{P}{N_{0}}$, for the band-limited case,
it only scales logarithmically with $\frac{P}{N_{0}}$.

In this paper, we improve on the converse bound of \eqref{eq: DPT band-limited}
using two different mechanisms. In the first, channel coding considerations,
as the ones used in the converse bound of \cite{Neri_estimation_DMC},
will be used to derive a converse bound to the problem at hand. In
the second method, we utilize the results of the unlimited-bandwidth
case from \cite{Cohn,Burnashev84,Burnashev85}, in a somewhat indirect
way, rather than revising the complicated bounding techniques used
to prove them. The general idea is to begin with a band-limited system,
and transform it, by some means, to a new system. We will then relate
the MP$\alpha$E exponent of the new system to the MP$\alpha$E exponent
of the original system, and use the converse bound of the unlimited-bandwidth
case, for the new system. This, in turn, will provide a converse bound
on the original, band-limited system. Two new bounds will be derived
from this general methodology. It turns out that none of the three
converse bounds mentioned above dominates the other two, and for each
of these bounds, there exists a region in the plane of the variables
$\alpha$ and SNR such that this is the best bound out of the three. 

To assess the tightness of the converse bounds, we will briefly discuss
also achievability bounds. Specifically, the achievability bound of
\cite{Neri_estimation_DMC} will be adapted to the AWGN channel, just
as the converse bound of \cite{Neri_estimation_DMC} was. We will
also speculate on a possible approach for improving this achievability
bound, based on \emph{unequal error protection} (even though, thus
far, we were not able to demonstrate that it actually improves). It
should be mentioned, that for this problem, converse bounds which
are based on other, well-known, estimation-theoretic lower bounds,
such as the Weiss-Weinstein bound \cite{Weiss_phd,Wiess_Weinstein},
have failed to provide stronger bounds, at least in the various ways
we have tried to harness them. 

The rest of the paper is organized as follows. In Section \ref{sec:System model},
the modulation-estimation problem is formulated, and known results
for the unlimited-bandwidth AWGN channel are reviewed. In Section
\ref{sec:Converse-Bounds}, the converse bound adapted from \cite{Neri_estimation_DMC}
is presented, and our main results, which are the two new converse
bounds on the MP$\alpha$E exponent. The achievability bound, also
adapted from \cite{Neri_estimation_DMC}, is discussed as well. In
Section \ref{sec:Results}, the various converse bounds are compared
to each other, as well as to the achievability bound. Numerical results
are displayed, and a systematic comparison between the bounds is made,
based on asymptotic SNR analysis.

\section{System Model and Background\label{sec:System model}}

Throughout the paper, real random variables will be denoted by capital
letters, and specific values they may take will be denoted by the
corresponding lower case letters. Random vectors and their realizations
will be denoted, respectively, by capital letters and the corresponding
lower case letters, both in the bold face font. Real random processes
will be denoted by capital letters with a time argument, and specific
sample paths will be denoted by the corresponding lower case letters.
For example, the random vector $\mathbf{X}=(X_{1},\ldots,X_{N})$,
($N$ positive integer) may take a specific vector value $\mathbf{x}=(x_{1},\ldots,x_{N})$,
and the random process $X(t)$ may have the sample path $x(t)$. The
probability of an event ${\cal E}$, for an underlying parameter $u\in[0,1)$,
will be denoted by $\P_{u}[{\cal E}]$, and the expectation operator
will be denoted by $\E_{u}[\cdot]$. The indicator for a set ${\cal A}$
will be denoted by $\I\{{\cal A}\}$. Logarithms and exponents will
be understood to be taken to the natural base. For the sake of brevity,
for large integers, we will ignore integer constraints throughout,
as they do not have any effect on the results. For example, we will
assume a blocklength $N=2WT$, rather than $N=\left\lceil 2WT\right\rceil $,
provided that $2WT\gg1$.

Let $u\in[0,1)$ be a parameter and consider the continuous-time AWGN
channel
\begin{equation}
y(t)=s(t,u)+z(t),\label{eq: Gaussian channel}
\end{equation}
where $s(t,u)$ and $y(t)$ are the channel input and output, respectively,
at time $t$, and $\{z(t)\}$ is a white Gaussian noise process with
two-sided spectral density $\frac{N_{0}}{2}$. 

\begin{figure}
\begin{centering}
\includegraphics{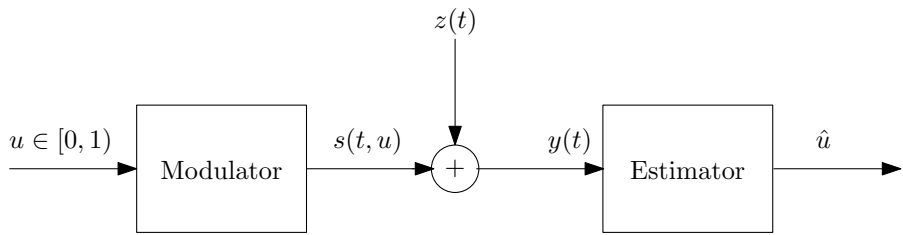}
\par\end{centering}

\protect\caption{A Modulation-estimation system.}

\end{figure}
A \emph{modulation-estimation} \emph{system} ${\cal S}_{T}$ of time
duration $T$ is defined by a \emph{modulator} and an \emph{estimator}.
The modulator maps%
\footnote{The mapping $u\to s(t,u)$ does not have to be necessarily injective
(one-to-one).%
} a parameter value $u$ to a signal $\{s(t,u),\;0\leq t\leq T\}$,
where $s(t,u)=0$ for $t<0$ and $t>T$, and where the mapping $u\to s(t,u)$
is assumed measurable. The estimator maps the received signal $\{y(t),\;0\le t\leq T\}$
to an estimated parameter, $\hat{u}$. The system ${\cal S}_{T}$
is \emph{power-limited} to $P$ if 
\begin{equation}
\frac{1}{T}\int_{0}^{T}s^{2}(t,u)\cdot\d t\leq P\label{eq: power limited}
\end{equation}
for all $u\in[0,1)$. The system is considered \emph{band-limited}
to $W$ if there exists an orthonormal basis of $N\dfn2WT$ functions
$\{\phi_{n}(t),\;0\leq t\leq T\}_{n=1}^{N}$, such that for all $u\in[0,1)$,
there exists a vector of coefficients, $\mathbf{s}(u)\dfn(s_{1}(u),\ldots,s_{N}(u))$,
such that
\begin{equation}
s(t,u)=\sum_{n=1}^{N}s_{n}(u)\cdot\phi_{n}(t),\quad0\leq t\leq T.\label{eq: band limited}
\end{equation}
Following a procedure similar to that of \cite[Section 2.1]{viterbi2009principles},
the continuous-time channel can be converted to an equivalent $N$-dimensional
channel. As discussed there, the projections
\begin{equation}
y_{n}\dfn\int_{0}^{T}y(t)\cdot\phi_{n}(t)\cdot\d t,\quad1\leq n\leq N,\label{eq: output signal projection}
\end{equation}
are \emph{sufficient statistics} for the estimation of $u$. We may
define the noise projections 
\begin{equation}
z_{n}\dfn\int_{0}^{T}z(t)\cdot\phi_{n}(t)\cdot\d t,\quad1\leq n\leq N,\label{eq: noise projections}
\end{equation}
and group the projections into vectors, $\mathbf{y}=(y_{1},\ldots,y_{N})$
and $\mathbf{z}=(z_{1},\ldots,z_{N})$, to obtain an equivalent vector
model 
\begin{equation}
\mathbf{y}\dfn\mathbf{s}(u)+\mathbf{z}.\label{eq: channel model discrete time}
\end{equation}
In this model, the power constraint is given by $\left\Vert \mathbf{s}(u)\right\Vert ^{2}\leq PT$,
but for the purpose of converse bounds, it can be assumed, without
loss of generality (w.l.o.g.), that the constraint is satisfied with
equality. Indeed, as was discussed in \cite[p. 249]{Burnashev85},
\cite[pp. 291-292]{wyner1973bound}, if $\left\Vert \mathbf{s}(u)\right\Vert ^{2}<PT$
for some $u$, then a single dummy coordinate can be appended to $\{\mathbf{s}(u)\}$,
which will make $\left\Vert \mathbf{s}(u)\right\Vert =PT$. For $N\gg1$,
this additional coordinate has a negligible effect on the time or
bandwidth of the signals, and, in fact, can be totally ignored by
the estimator. Regarding the noise, as the projection in \eqref{eq: noise projections}
is performed on an orthogonal set, the resulting projections are independent,
and thus $\mathbf{Z}\sim{\cal N}(\mathbf{0},\frac{N_{0}}{2}\cdot I_{N})$,
where $I_{N}$ is the identity matrix of dimension $N$. The estimator,
based on the channel \eqref{eq: channel model discrete time}, can
then be denoted as a function of $\mathbf{y}$, i.e., $\hat{u}(\mathbf{y})$
rather than $\hat{u}\{y(t),\;0\le t\leq T\}$ for \eqref{eq: Gaussian channel}. 

At this point, a justification for adopting \eqref{eq: channel model discrete time}
as a proper model for a physically band-limited channel is required.
The correspondence between the continuous-time model \eqref{eq: Gaussian channel}
and the discrete-frequency model \eqref{eq: channel model discrete time}
is a delicate, yet a mature subject. In short, signals cannot be both
strictly time-limited and strictly band-limited. Thus, the basis functions
$\{\phi_{n}(t),\;0\leq t\leq T\}_{n=1}^{N}$ are chosen to span the
linear space of signals of duration $T$ exactly, and a bandwidth
of \emph{approximately} $W$.%
\footnote{These basis functions are known as \emph{prolate spheroidal functions.}%
} If $N=2WT\gg1$, the proximity of the real bandwidth to $W$ can
be made arbitrarily sharp. A detailed discussion can be found in \cite{On_bandwidth,wyner1966capacity},
and \cite[Chapter 8]{gallager1968information}. 

For $\alpha>0$ (not necessarily integer), the \emph{mean power-$\alpha$
error (MP$\alpha$E)} of ${\cal S}_{T}$ is defined as
\begin{equation}
e_{\alpha}({\cal S}_{T})\dfn\sup_{u\in[0,1)}\E_{u}\left\{ \left|\hat{u}(\mathbf{Y})-u\right|^{\alpha}\right\} ,\label{eq: mean-alpha-error}
\end{equation}
where $\mathbf{Y}$ is the random counterpart of $\mathbf{y}$. As
we shall see, $e_{\alpha}({\cal S}_{T})$ can be made exponentially
decreasing with $T$, and so, it is natural to ask what is the fastest
possible exponential rate of decrease. Specifically, we say that $E$
is an achievable MP$\alpha$E exponent if there exists a family $\{{\cal S}_{T}\}$
of modulation-estimation systems, parametrized by $T$, such that
\[
\limsup_{T\to\infty}\left[-\frac{1}{T}\cdot\log e_{\alpha}({\cal S}_{T})\right]\geq E.
\]
The objective of the paper is to derive converse bounds on $E_{\alpha}(\nicefrac{P}{N_{o}},W)$,
which is defined as the largest achievable MP$\alpha$E exponent,
for a given power constraint $P$, bandwidth constraint $W$, and
noise spectral density $\frac{N_{0}}{2}$. Let us define the SNR as
$\Gamma\dfn\frac{P}{N_{0}W}$. Noting that power constraint on the
input to the channel \eqref{eq: channel model discrete time} can
be written as 
\[
\frac{\left\Vert \mathbf{s}(u)\right\Vert ^{2}}{N}\leq\frac{PT}{N}=\frac{PT}{2WT}=\frac{P}{2W}.
\]
Scaling $\mathbf{y}$ by $\sqrt{\frac{2}{N_{0}}}$, we get an equivalent
channel
\begin{equation}
\tilde{\mathbf{y}}\dfn\tilde{\mathbf{s}}(u)+\tilde{\mathbf{z}},\label{eq: channel model discrete time equivalent}
\end{equation}
with a power input constraint
\[
\frac{\left\Vert \tilde{\mathbf{s}}(u)\right\Vert ^{2}}{N}\leq\Gamma,
\]
and $\tilde{\mathbf{z}}\sim{\cal N}(\mathbf{0},I_{N}).$ Note that
the dimension of the channel \eqref{eq: channel model discrete time}
and \eqref{eq: channel model discrete time equivalent} is given by
$N=2WT$. Since the properties of the channel \eqref{eq: channel model discrete time equivalent}
depend on $W$ and $T$ only via their product $WT$, for a fixed
SNR, scaling the bandwidth $W$ by a factor $a>0$ has the same effect
as scaling $T$ by $a$ instead.%
\footnote{Of course, to keep the SNR fixed, the power should be changed to $a\cdot P$.%
} Thus, the MP$\alpha$E exponent will always have the form 
\begin{equation}
E_{\alpha}\left(\frac{P}{N_{0}},W\right)=W\cdot F_{\alpha}(\Gamma),\label{eq: exponent per unit bandwidth}
\end{equation}
where $F_{\alpha}(\Gamma)$ is a certain function. The same comment
applies to the converse and achievability bounds that will be encountered
along this work. So, henceforth, we will be interested in the \emph{MP$\alpha$E
exponent per unit bandwidth} $F_{\alpha}(\Gamma)$. Note that the
resulting MP$\alpha$E has the exponential form $\exp[-TW\cdot F_{\alpha}(\Gamma)]=\exp[-\frac{N}{2}\cdot F_{\alpha}(\Gamma)]$.
Most of the time, it will be convenient to carry out the exponent
analysis in the discrete domain, and then finally, translate the result
to the exponent \eqref{eq: exponent per unit bandwidth}, simply by
doubling the exponent. 

To review the known converse bounds for the unlimited-bandwidth case,
we begin by formulating the appropriate scaling of their MP$\alpha$E
exponent. Writing \eqref{eq: exponent per unit bandwidth} as 
\begin{equation}
E_{\alpha}\left(\frac{P}{N_{0}},W\right)=W\cdot F_{\alpha}(\Gamma)=W\cdot\frac{F_{\alpha}(\Gamma)}{\Gamma}\cdot\Gamma\label{eq: infinite bandwith exponent def}
\end{equation}
and noting that as $W\to\infty$ then $\Gamma\to0$, we can define
\emph{unlimited-bandwidth MP$\alpha$E exponent }as
\[
\gamma_{\alpha}\dfn\lim_{\Gamma\to0}\frac{F_{\alpha}(\Gamma)}{\Gamma}.
\]
Thus, for $W\to\infty$, \eqref{eq: infinite bandwith exponent def}
has the same form as \eqref{eq: exponent per unit bandwidth}, with
the exponent per unit bandwidth being a \emph{linear} function of
the SNR, as $\gamma_{\alpha}\Gamma$. By contrast, as we shall see
in Section \ref{sec:Converse-Bounds}, and as was mentioned earlier,
for band-limited signals, $F_{\alpha}(\Gamma)$ scales logarithmically
with $\Gamma$. 

The value of $\gamma_{\alpha}$ was bounded by Cohn \cite{Cohn},
and later on by Burnashev \cite{Burnashev84,Burnashev85}. The best
known converse bound is given by \cite[Theorem 2]{Burnashev84}, \cite[Theorem 3]{Burnashev85}
\footnote{\label{fn:To-translate-Burnashev's}To translate Burnashev's results
to our defintions, the value of the exponent in \cite{Burnashev84,Burnashev85}
should be doubled. In the notation of \cite{Burnashev84,Burnashev85},
the MP$\alpha$E is an exponential function of the energy per noise
spectral density, and has the form $\exp(-\gamma_{\alpha}\cdot A)$
where $A=\frac{PT}{\nicefrac{N_{0}}{2}}$ (compare with \eqref{eq: infinite bandwith exponent def}).%
}
\begin{equation}
\gamma_{\alpha}\leq\begin{cases}
\frac{1}{(1+\alpha)}\min\left\{ \alpha,\psi(\alpha)\right\} , & 0<\alpha\leq\alpha_{0}\\
\frac{\alpha}{2(1+\alpha)}\left[1+\frac{\alpha+5-4\sqrt{\alpha+1}}{3\alpha+1}\right], & \alpha_{0}\leq\alpha\leq2\\
\frac{\alpha}{2(1+\alpha)}, & \alpha\geq2
\end{cases},\label{eq: Burnashev bound}
\end{equation}
where $\alpha_{0}$ is the unique root of the equation $\alpha^{2}-(\alpha-1)\sqrt{\alpha+1}-2=0$
($\alpha_{0}\approx1.5875$) and
\[
\psi(\alpha)\dfn1+\alpha-\max_{q\geq\nicefrac{1}{2}}\left[2\alpha q+4q\sqrt{(1-q)q(1+\alpha)}-q^{2}(3\alpha+1)\right].
\]
In fact, for $\alpha\geq2$ this is the exact value of $\gamma_{\alpha}$
as there are schemes that achieve it (\cite[Theorem 1]{Burnashev84}
and c.f. Remark \ref{rem: channel coding achievable for unlimited bandwidth channels}).

\section{Exponential Bounds on the MP$\alpha$E \label{sec:Converse-Bounds}}

In this section, we present three new converse bounds on the MP$\alpha$E
exponent. The first bound is an adaptation of the converse bound of
\cite{Neri_estimation_DMC}, originally derived for modulation-estimation
over discrete memoryless channels, and this bound will be termed the\emph{
channel coding converse bound}. The proof idea is to relate the MP$\alpha$E
exponent of a modulation-estimation system to the error exponent of
an optimal channel code (reliability function). Since the error exponent
of channel codes is lower bounded by the sphere-packing exponent (or
any other upper bound on the reliability function), a converse bound
on the MP$\alpha$E exponent is obtained. 

We then derive two additional converse bounds by converting the unlimited-bandwidth
bound to the band-limited case, and these are the main results of
this paper. An appealing property of these two bounds is that their
proof is only based on the value of the unlimited-bandwidth converse
bound, and not on the way it was proved. Consequently, there is no
need to repeat the intricate proofs of the unlimited-bandwidth bound
in order to derive the new bounds. Further, any future improvement
of the bound \eqref{eq: Burnashev bound} will immediately lend itself
to a corresponding improvement of our band-limited bounds. 

The first bound of this type will be referred to as the \emph{spherical
cap bound,} and its derivation is based on the following idea. The
signal vectors of any band-limited system reside on the surface of
a sphere of radius $\sqrt{PT}$, centered at the origin. For any given
angle, there exists a spherical cap in the surface of this sphere,
such that the signal vectors confined to this spherical cap pertain
to a significant portion (depending on the angle) of the parameter
domain $[0,1)$. Then, a new modulation-estimation system can be constructed,
which is based only on signals which lie in this spherical cap. While
this new system is still band-limited, its exponent must obviously
obey the unlimited converse bound. This in turn leads to a converse
bound on the original system, whose tightest value is obtained by
optimization of the aforementioned angle.

The second bound will be referred to as the \emph{spectrum replication
bound,} and it is based on creating many replicas of the signal set
of a given band-limited modulation system in higher frequency bands.
This results in a new system, where the value of the modulated parameter
determines which of the frequency bands will be active, and which
signal will be transmitted within the band. As this new system has
much larger bandwidth, it is proper to bound its MP$\alpha$E exponent
by the unlimited-bandwidth bound, which in turn, leads to a bound
on the original system, whose MP$\alpha$E exponent is easily related
to that of the duplicated wideband system. 

In the rest of the section, we will outline the derivation of each
of the bounds in somewhat more detail, and then formally state it.
The formal proofs of the spherical cap bound and the spectrum replication
bound will be relegated to Appendix \ref{sec:Proof-of-Converse-Bounds}.
Then, we will briefly discuss also the weaknesses of the various bounds.
Finally, we will discuss and state an achievability bound, which is
also based on an analogous bound from \cite{Neri_estimation_DMC},
and then discuss its possible weaknesses, along with some speculations
on how it might be strengthened.

The proof of the channel coding converse bound begins by employing
Chebyshev\textquoteright s inequality, to link the MP$\alpha$E and
the large deviations performance of the system as follows 
\begin{equation}
\E_{u}\left\{ \left|\hat{u}(\mathbf{Y})-u\right|^{\alpha}\right\} \geq\Delta^{\alpha}\cdot\P_{u}\left\{ \left|\hat{u}(\mathbf{Y})-u\right|>\Delta\right\} .\label{eq: Chebyshev on mean-alpha-error}
\end{equation}
Then, an arbitrary rate $R$ is chosen and $\Delta=\exp(-NR)$ is
set. In \cite[Theorem 1]{Neri_modulation_estimation}, it is shown
that if there exists a modulation-estimation system such that $\P_{u}\left\{ \left|\hat{u}(\mathbf{Y})-u\right|>e^{-NR}\right\} $
decays with some exponent $E(R)$, then an ordinary channel code of
rate $R$ can be constructed which achieves the same exponent. Thus,
as $E(R)$ cannot be larger than the reliability function of channel
coding, it follows from \eqref{eq: Chebyshev on mean-alpha-error}
that the MP$\alpha$E exponent cannot be larger than $E(R)+\alpha R$.
Finally, the best bound is obtained by optimizing over the rate $R$,
to yield $\min_{R\geq0}\left[E(R)+\alpha R\right]$. 

To state the bound more explicitly, let us define Gallager's random
coding function \cite[p. 339, eq. (7.4.24)]{gallager1968information}
\begin{equation}
E_{0}(\rho,\Gamma)\dfn\frac{1}{2}\left[(1-\beta_{0})(1+\rho)+\Gamma+\log\left(\beta_{0}-\frac{\Gamma}{1+\rho}\right)+\rho\log(\beta_{0})\right],\label{eq: E0}
\end{equation}
where \cite[p. 339, eq. (7.4.28)]{gallager1968information} 
\begin{equation}
\beta_{0}\dfn\frac{1}{2}\left(1+\frac{\Gamma}{1+\rho}\right)\left[1+\sqrt{1-\frac{4\Gamma\rho}{(1+\rho+\Gamma)^{2}}}\right],\label{eq: beta_0}
\end{equation}
and Gallager's expurgated function \cite[p. 341, eq. (7.4.43)]{gallager1968information}
\begin{equation}
E_{\st[x]}(\rho,\Gamma)\dfn(1-\beta_{\st[x]})\rho+\frac{\Gamma}{2}+\frac{\rho}{2}\log\left[\beta_{\st[x]}\left(\beta_{\st[x]}-\frac{\Gamma}{2\rho}\right)\right],\label{eq: Ex}
\end{equation}
where \cite[p. 342, eq. (7.4.45)]{gallager1968information}
\begin{equation}
\beta_{\st[x]}\dfn\frac{1}{2}+\frac{\Gamma}{4\rho}+\frac{1}{2}\sqrt{1+\frac{\Gamma^{2}}{4\rho^{2}}}.\label{eq: beta_x}
\end{equation}
It should be remarked that for the converse bound on the MP$\alpha$E
exponent, Gallager's random coding exponent is used only at rates
for which it equals to the reliability function of channel codes,
namely, where it coincides with the sphere-packing exponent. In addition,
it is well known that the channel coding reliability function at zero
communication rate is equal to the expurgated exponent, which in turn
is given by 
\[
\lim_{\rho\to\infty}E_{\st[x]}(\rho,\Gamma)=\frac{\Gamma}{4}.
\]
We now have the following Proposition.
\begin{prop}[Channel coding converse bound]
\label{prop:Channel coding upper bound}The MP$\alpha$E exponent
per unit bandwidth is upper bounded as
\begin{equation}
F_{\alpha}(\Gamma)\leq\min\left\{ 2E_{0}(\alpha,\Gamma),\gamma_{\alpha}\Gamma\right\} .\label{eq: Channel Coding Upper Bound}
\end{equation}
\end{prop}
\begin{IEEEproof}
Using the same proof as in \cite[Theorem 1, Appendix A]{Neri_estimation_DMC}
and outlined above, we have that 
\[
F_{\alpha}(\Gamma)\leq2\cdot\min\left\{ E_{0}(\alpha,\Gamma),\frac{\Gamma}{4}\right\} .
\]
For the band-limited AWGN channel, we may also add to the minimization
the unlimited-bandwidth bound, and so 
\begin{equation}
F_{\alpha}(\Gamma)\leq\min\left\{ 2E_{0}(\alpha,\Gamma),\frac{\Gamma}{2},\gamma_{\alpha}\Gamma\right\} .\label{eq: channel coding converse bound proof - minimization of three term}
\end{equation}
Now, by definition, $\gamma_{\alpha}$ is non-decreasing with $\alpha$,
and from \eqref{eq: Burnashev bound} $\lim_{\alpha\to\infty}\gamma_{\alpha}=\frac{1}{2}$.
Thus, $\gamma_{\alpha}\leq\frac{1}{2}$ and so $\frac{\Gamma}{2}$
never dominates the minimization in \eqref{eq: channel coding converse bound proof - minimization of three term}.
\end{IEEEproof}
Note that in the channel coding converse bound, the variable $\rho$
of Gallager's random coding function is set to $\alpha$, and can
be larger than $1$, because the function $E_{0}(\alpha,\Gamma)$
actually arises from the sphere-packing exponent, for which $\rho$
is positive and not limited to $[0,1]$. 

\begin{figure}
\centering{}\includegraphics[scale=0.8]{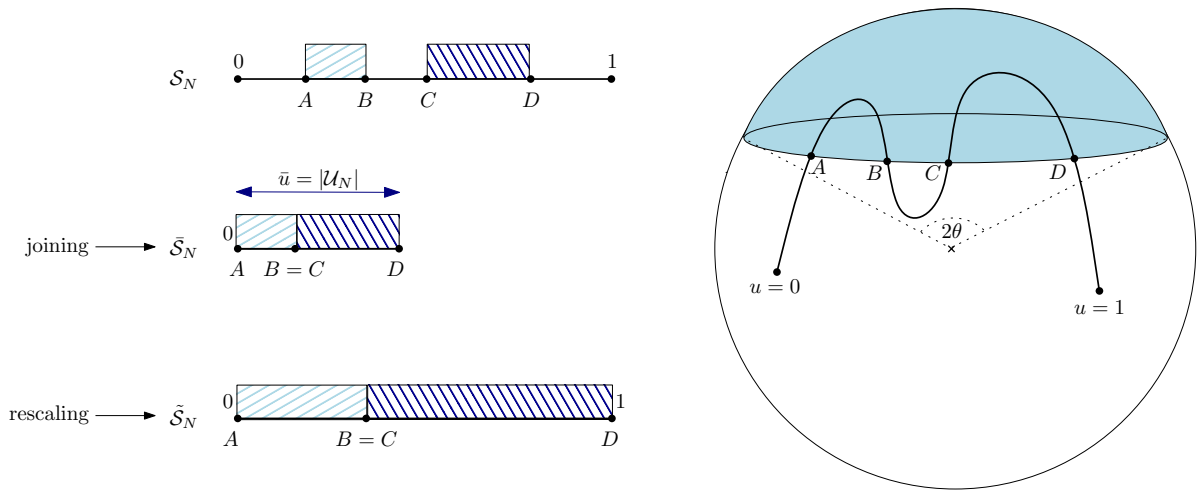}\protect\caption{Illustration of the mechanism of the spherical cap bound.\label{fig: spherical cap proof}}
\end{figure}

\begin{figure}
\centering{}\includegraphics[scale=1.2]{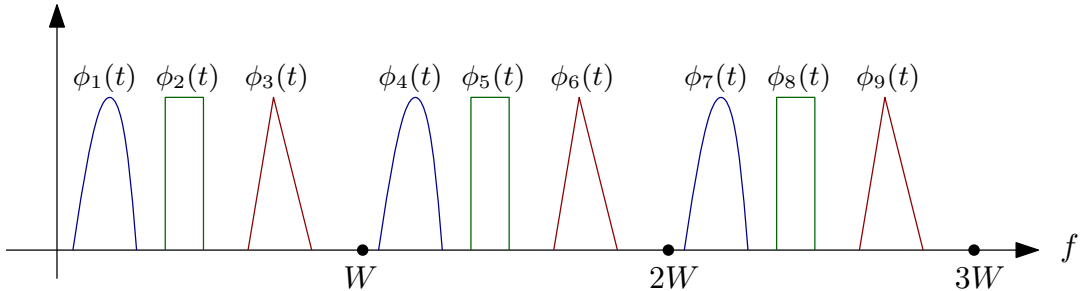}\protect\caption{The system ${\cal S}_{T}$ is band-limited to $W$, and uses a linear
combination of the orthonormal basis $\{\phi_{n}(t)\}_{n=1}^{N}$
to modulate the parameter $u$, where, here $N=3$. The orthonormal
basis $\{\phi_{n}(t)\}_{n=1}^{3}$ is duplicated, using a frequency
translation to the bands $[W,2W)$ and $[2W,3W)$. The new system
$\tilde{{\cal S}}_{T}$, shown here with $M=3$, modulates the parameter
$v$ by first choosing a frequency band $m\in\{0,1,2\}$, and then
modulates $\{\phi_{n}(t)\}_{n=mM+1}^{mM+N}$, just as ${\cal S}_{T}$
modulates $\{\phi_{n}(t)\}_{n=1}^{N}$.\label{fig: superposition spectrum proof}}
\end{figure}

The outline of the derivation of the spherical cap bound is as follows.
With some abuse of notation, the system ${\cal S}_{N}$ will be identified
with the projection vectors of its signal set, ${\cal S}_{N}\dfn\left\{ \mathbf{s}(u):\; u\in[0,1)\right\} $,
and its MP$\alpha$E will be denoted by $e_{\alpha}({\cal S}_{N})$.
We begin with an arbitrary band-limited system ${\cal S}_{N}$. As
can be seen in Fig. \ref{fig: spherical cap proof}, 
 only part of the locus, created by the signals in ${\cal S}_{N}$,
is contained in a given spherical cap of angle $2\theta$. If we focus
only on the subset of parameters values pertaining to signals within
the spherical cap, and join these subsets to the left (see Fig. \ref{fig: spherical cap proof}),
we get a new system $\overline{{\cal S}}_{N}$ which modulates parameters
in $[0,\overline{u})$ for some $\overline{u}\leq1$, and uses the
signals within the spherical cap only. If we then rescale the interval
$[0,\overline{u})$ back to $[0,1)$ (while still using only signals
within the same spherical cap), we get a new system $\tilde{{\cal S}}_{N}$,
for parameters in $[0,1)$. The MP$\alpha$Es of the various systems
${\cal S}_{N}$, $\overline{{\cal S}}_{N}$ and $\tilde{{\cal S}}_{N}$
obey a simple relationship, and thus any bound on the MP$\alpha$E
of $\tilde{{\cal S}}_{N}$ implies a bound on the MP$\alpha$E of
${\cal S}_{N}$. Specifically, using the unlimited-bandwidth converse
bound \eqref{eq: Burnashev bound} on the MP$\alpha$E exponent of
$\tilde{{\cal S}}_{N}$ (even though it is a band-limited system)
leads to the spherical cap bound. A key point in the proof is a measuring
argument similar to \cite[pp. 293-294]{wyner1973bound}, which is
used to prove the existence of a spherical cap which contains a significant
portion of the signal set locus. Finally, as the angle $\theta$ of
the spherical cap was arbitrary, it is optimized to obtain the tightest
bound. The following theorem is then obtained.
\begin{thm}[Spherical cap bound]
\label{thm:Spherical cap bound}The MP$\alpha$E exponent per unit
bandwidth is upper bounded as
\begin{equation}
F_{\alpha}(\Gamma)\leq\begin{cases}
\gamma_{\alpha}\Gamma, & \Gamma<\frac{\alpha}{\gamma_{\alpha}}\\
\alpha\left[\log\left(\frac{\gamma_{\alpha}\Gamma}{\alpha}\right)+1\right], & \Gamma\geq\frac{\alpha}{\gamma_{\alpha}}
\end{cases}.\label{eq: Spherical Cap Bound}
\end{equation}

\end{thm}
Next, we outline the derivation of the spectrum replication bound.
The proof relies on the idea of superimposing a frequency position
modulation over a system ${\cal S}_{T}$ for bandwidth $W$. Suppose
that we have a system ${\cal S}_{T}$ whose signals are band-limited
to $[0,W)$. Imagine that we duplicate its signal set by a simple
frequency shifts, from the frequency band $[0,W)$ to all the frequency
bands $[mW,(m+1)W)$ for $0\leq m\leq M-1$, where $M$ is integer,
thus obtaining a new signal set for a system $\tilde{{\cal S}}_{T}$.
Now, a specific signal in the new signal set is specified by two components
of the parameter: (i) the frequency band index $m$, and (ii) the
signal within the band, which is nothing but a frequency translation
of a signal from ${\cal S}_{T}$. The spectrum of the signals of ${\cal S}_{T}$
and $\tilde{{\cal S}}_{T}$ is illustrated in Fig. \ref{fig: superposition spectrum proof}.
Accordingly, we can construct a modulation-estimation system $\tilde{{\cal S}}_{T}$
which modulates both parameters. 

Specifically, let the newly constructed system be denoted by $\tilde{{\cal S}}_{T}$.
The parameter at the input of this system is first uniformly quantized
to $M$ values, and then the quantization error, after a proper scaling
to $[0,1)$, is used as an input to the original system ${\cal S}_{T}$.
The signal chosen from ${\cal S}_{T}$ is then modulated to one of
$M$ possible non-overlapping frequency bands according to the quantized
value of the parameter, and then transmitted over the channel. 

At the receiver, first the active frequency band is decoded using
a \emph{non-coherent} decoder, and the quantized part is estimated.
Then, the signal is demodulated to baseband (assuming a correct decoding
at the first stage), and the estimator of ${\cal S}_{T}$ is used
to estimate the quantization error. Afterwards, an estimation of the
parameter is obtained using both the decoded quantized value and the
estimation of the quantization error. 

Now, on the one hand, the MP$\alpha$E exponent of the new system
$\tilde{{\cal S}}_{T}$ can be lower bounded by an expression which
depends on the MP$\alpha$E exponent of ${\cal S}_{T}$, i.e., $F_{\alpha}(\Gamma)$,
and the probability of correct modulation frequency decoding. On the
other hand, the signals of $\tilde{{\cal S}}_{T}$ occupy the frequency
band $[0,MW)$, and if $M\gg1$, %
\footnote{As we shall see, $M$ is in fact chosen to exponentially increasing
with $T$. %
} these signals have a much larger bandwidth than the original system.
Thus, it is proper to upper bound the MP$\alpha$E exponent of the
new system $\tilde{{\cal S}}_{T}$ by the unlimited-bandwidth bound
\eqref{eq: Burnashev bound}. Using these relations, a bound on $F_{\alpha}(\Gamma)$
can be readily obtained. 

To state the spectrum replication bound, we need the following definitions.
For $\rho\in[0,1]$ define%
\footnote{This function plays the role of Gallager's $E_{0}(\rho)$ function
in the random coding exponent for ordinary channel coding \cite[Section 5.6]{gallager1968information}.%
} 
\begin{equation}
\Phi(\rho,\Gamma)\dfn\rho[\eta-1-\log\eta]+\eta+\Gamma+\log\left[\frac{\sqrt{4\eta\Gamma+1}+1}{2\eta}\right]-\sqrt{4\eta\Gamma+1},\label{eq: Phi definition}
\end{equation}
where
\begin{equation}
\eta=\frac{\Gamma+\sqrt{\Gamma^{2}-4(\rho^{2}+1)(\rho+1)^{2}}}{2(\rho+1)^{2}},\label{eq: eta star}
\end{equation}
and also define 
\begin{equation}
\Lambda_{\alpha}(\Gamma)\dfn\sup_{0<\rho\leq1}\left\{ \frac{\Phi(\rho,\Gamma)-\gamma_{\alpha}^ {}\Gamma}{\rho}\right\} .\label{eq: Lambda defintion}
\end{equation}

\begin{thm}[Spectrum replication bound]
\label{thm: spectrum replication bound}The MP$\alpha$E exponent
per unit bandwidth is upper bounded as
\begin{equation}
F_{\alpha}(\Gamma)\leq\gamma_{\alpha}\Gamma-\left[\alpha\Lambda_{\alpha}(\Gamma)\right]_{+}\label{eq: Superposition Bound}
\end{equation}
where $[t]_{+}\dfn\max\{t,0\}$.
\end{thm}
It is evident that both bounds of \eqref{eq: Spherical Cap Bound}
and \eqref{eq: Superposition Bound} are monotonically increasing
with $\gamma_{\alpha}$. Thus, in the range where the true value of
$\gamma_{\alpha}$ is not known ($0<\alpha<2$), any upper bound on
$\gamma_{\alpha}$ can be used, in particular, the bound \eqref{eq: Burnashev bound}.

As we shall see in Section \ref{sec:Results}, all the three converse
bounds mentioned above, as far as we know, are the best available,
at least for some $\alpha$ and $\Gamma$. However, for the sake of
potential future improvement of these bounds, it is insightful to
point out also their weaknesses. As discussed in \cite[p. 839, footnote 6]{Neri_estimation_DMC},
the weakness of the channel coding converse bound does not stem from
the use of Chebyshev\textquoteright s inequality, but from the fact
that there is no apparent single estimator which minimizes $\P_{u}\left\{ \left|\hat{u}(\mathbf{Y})-u\right|>e^{-NR}\right\} $,
uniformly for all $R$. The spherical cap bound suffers from the fact
that an unlimited-bandwidth bound is used as a converse bound within
the cap. The spectrum replication bound has the weakness that it is
based on analyzing a two-stage estimator, which first decodes the
frequency band, and then uses the signal in this band to estimate
the parameter. Furthermore, in the first step, the frequency band
is decoded using a sub-optimal, non-coherent decoder. Nonetheless,
the above weaknesses are the result of compromises made to make the
analysis reasonably tractable, and, as said, give non-trivial results.

We conclude this section with an achievability bound. The idea is
to use a separation-based scheme, which first uniformly quantizes
the parameter to $\exp(NR)$ points, for some $R>0$. Then, it maps
the quantized parameter to a codeword from an ordinary channel code,
which achieves the reliability function, $E(R,\Gamma)$. At the receiver,
the maximum likelihood channel decoder is used to decode the transmitted
codeword, and the estimated parameter is defined as the midpoint of
the quantization interval of the decoded codeword. Note that increasing
the rate $R$, reduces the quantization error, but increases the probability
of decoding error and vice-versa. Thus, the rate is optimized in order
to maximize the MP$\alpha$E exponent. 

The derivation of this bound is a straightforward extension of \cite{Neri_estimation_DMC}.
We denote by $E(R,\Gamma)$ the reliability function of the AWGN channel
\eqref{eq: channel model discrete time} with SNR $\Gamma$, i.e.,
the maximal achievable error exponent for sequence of codes of rate
$R$. As is well known, it can be assumed that the reliability function
is for the maximal error probability over all codewords. We will use
the definitions in \eqref{eq: E0} and \eqref{eq: Ex}.
\begin{prop}[Achievability bound]
\label{prop:Channel coding lower bound}The MP$\alpha$E exponent
per unit bandwidth is lower bounded as
\begin{align}
F_{\alpha}(\Gamma) & \geq2\cdot\max_{R\geq0}\min\left\{ E(R,\Gamma),\alpha R\right\} \label{eq: Channel coding lower bound reliability}\\
 & \geq2\cdot\max\left\{ \sup_{0\leq\rho\leq1}\frac{\alpha E_{0}(\rho,\Gamma)}{\rho+\alpha},\sup_{\rho\geq1}\frac{\alpha E_{\st[x]}(\rho,\Gamma)}{\rho+\alpha}\right\} .\label{eq: Channel Coding Lower Bound}
\end{align}
\end{prop}
\begin{rem}
\label{rem: channel coding achievable for unlimited bandwidth channels}An
achievable bound for the unlimited-bandwidth AWGN channel can be proved
similarly to Proposition \ref{prop:Channel coding lower bound}. In
this case, the reliability function $E(R,\Gamma)$ in \eqref{eq: Channel coding lower bound reliability}
is known exactly for all rates. With a slight change of arguments,
it is given by \cite{wyner1967probability} 
\[
E(R,C)=\begin{cases}
\frac{C}{2}-R, & 0\leq R\leq\frac{C}{4}\\
(\sqrt{C}-\sqrt{R})^{2}, & \frac{C}{4}\leq R\leq C
\end{cases},
\]
where $C=\frac{P}{N_{0}}$. Since $\alpha R$ ($E(R,C)$) is an increasing
(respectively, decreasing) function of $R$, when $\alpha\geq2$,
the solution of $2\cdot\max_{R\geq0}\min\left\{ E(R,C),\alpha R\right\} $
is obtained at $R=\frac{C}{2(\alpha+1)}=\frac{1}{2(\alpha+1)}\cdot\frac{P}{N_{0}}$.
This proves the tightness of \eqref{eq: Burnashev bound} for $\alpha\geq2$. 
\end{rem}

\begin{rem}
It was shown in \cite{Neri_estimation_DMC} that this bound is tight
in the extreme cases of $\alpha\to0$ and $\alpha\to\infty$. This
is indeed plausible since when $\alpha\to0$ the error $|\hat{u}-u|^{\alpha}$
behaves like a ``zero-one'' loss function, in the sense that large
errors do not incur more penalty than small errors. Thus, in this
case, the quantization error dominates the MP$\alpha$E, and the rate
is maximized, i.e. chosen to be the channel capacity. A similar situation
occurs when $\alpha\to\infty$, but that in this case, the error $|\hat{u}-u|^{\alpha}$
tends to be a ``zero-infinity'' loss function. Large errors still
do not penalize more than small errors, but any error event causes
a catastrophically large penalty. Thus, in this case, the decoding
error dominates the MP$\alpha$E, and the rate is minimized in order
to maximize the decoding reliability, i.e. chosen to be zero. It should
be stressed, however, that the achievability and converse bound are
tight for a given $\Gamma$, as $\alpha\to0$ and $\alpha\to\infty$,
but may not be the best bounds for a \emph{given} $0<\alpha<\infty$. 
\end{rem}
An apparent weakness of the achievability bound is that it is derived
from analyzing a separation-based system, which means that the mapping
between one of the $M$ possible quantized parameter values and the
$M$ signal is arbitrary. A better system should choose this mapping
such that nearby (quantized) parameter values will be mapped to nearby
signals. In this case, a decoding error will typically cause only
a small error in the parameter value. In other words, if one maps
the quantized parameter value into bits, an unequal error protection
scheme should be used to communicate these bits \cite{BNL09}, with
larger reliability for the most significant bits than for the least
significant bits.

Typically, such a scheme uses an \emph{hierarchical channel code}
(also called \emph{superposition coding}) \cite{Bergmans_broadcast},
just like the one used, e.g., for the \emph{broadcast channel} \cite[Chapter 5]{el2011network}.
Each codeword, in this case, is given by the sum of a `cloud' codeword
and a `satellite' codeword,%
\footnote{In a two-users degraded broadcast channel, the cloud codeword carries
the message to be decoded by both users, while the satellite codeword
carries the private message, intended for the strong user only. %
} where the most significant bits determine the cloud codeword, and
the least significant bits determine the satellite codeword. The advantage
of such a system is that pairs of signals pertaining to nearby parameter
values belong to the same `cloud', whereas pairs of signals that are
associated with distant parameter values are allowed to belong to
different clouds. Thus, when a satellite decoding error occurs, this
results in only an error in the refined part of the quantized parameter.
Since the cloud centers have a rate lower than the entire codebook,
the decoding error probability of the cloud centers can be significantly
reduced, and overall, lead to a better MP$\alpha$E exponent. It can
also be noticed that a scheme in the same spirit was used in the spectrum
replication bound (Theorem \ref{thm: spectrum replication bound}),
as a method to prove a converse bound on the exponent.

Unfortunately, despite a considerable effort in this direction, we
were not able to find a concrete bound which improves the achievability
bound. It seems that the problem is that strong bounds on the MP$\alpha$E
can be obtained only by analyzing the optimal cloud decoder (as, e.g.,
in \cite{Kaspi_exponent_broadcast}), and not a decoder which treats
the interference from the satellite as noise (as, e.g., in \cite{Weng_error_exponent_broadcast}).
Especially, it seems that expurgated bounds for optimal cloud decoding
are most useful for the problem of bounding the MP$\alpha$E. However,
the best expurgated bound we are aware of was not sufficiently strong
to improve the achievability bound on the MP$\alpha$E.

\section{Results and Comparison among the Bounds \label{sec:Results}}

\begin{figure}
\centering{}\includegraphics[scale=0.4]{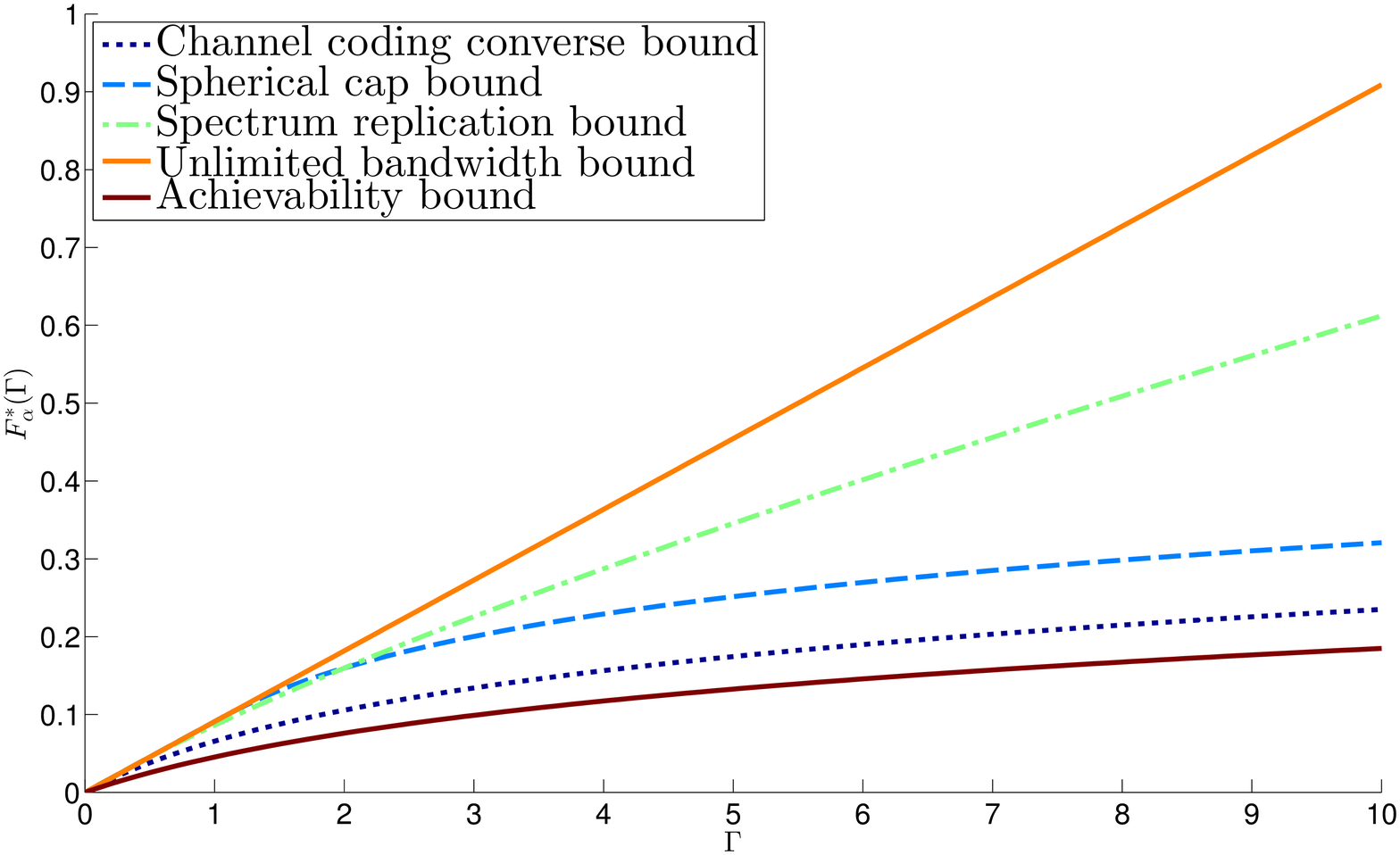}\protect\caption{Various bounds on $F_{\alpha}(\Gamma)$ for $\alpha=0.1$. \label{fig:alpha=00003D0.1}}
\end{figure}

\begin{figure}
\centering{}\includegraphics[scale=0.4]{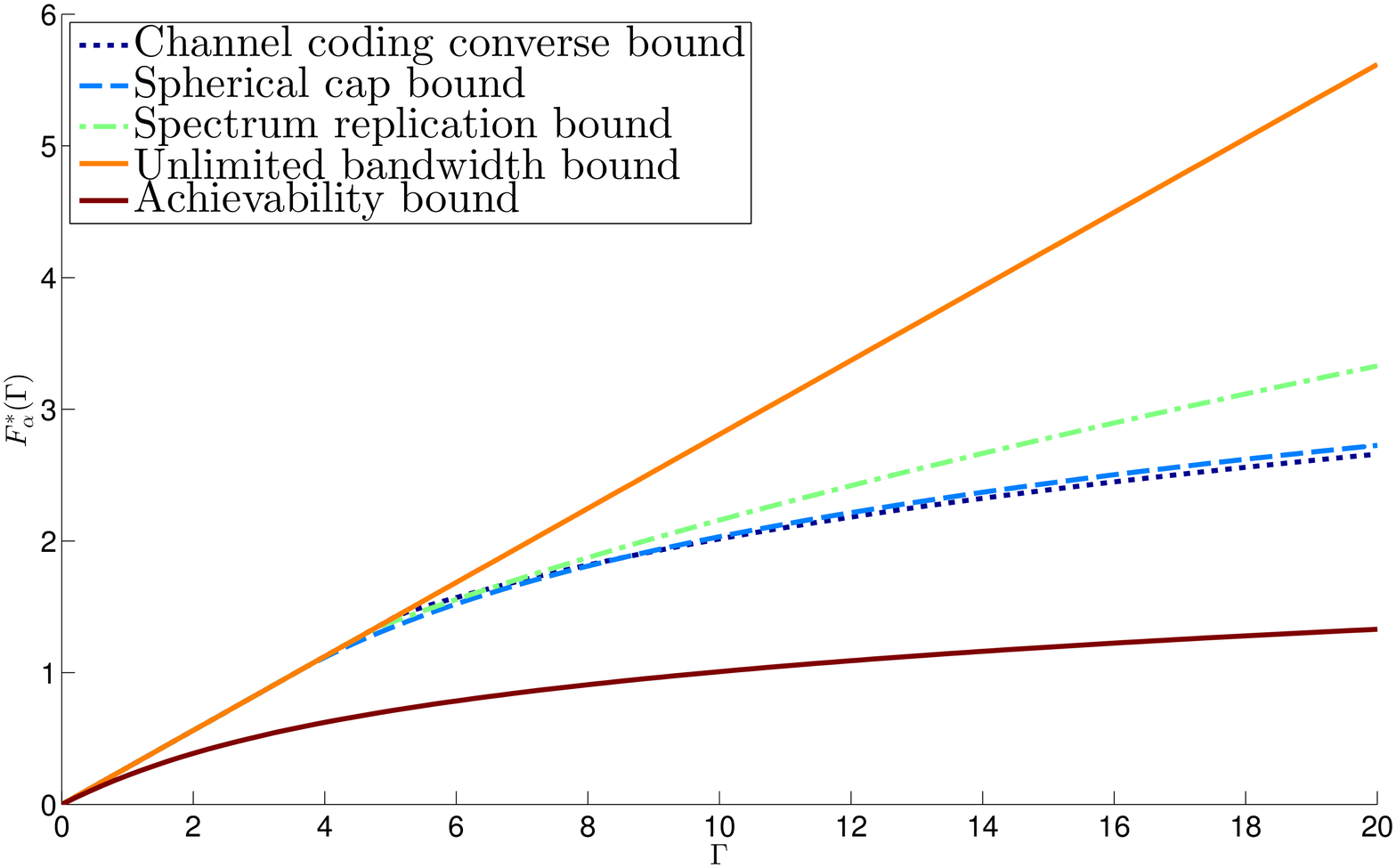}\protect\caption{Various bounds on $F_{\alpha}(\Gamma)$ for $\alpha=1$. \label{fig:alpha=00003D1}}
\end{figure}

\begin{figure}
\centering{}\includegraphics[scale=0.4]{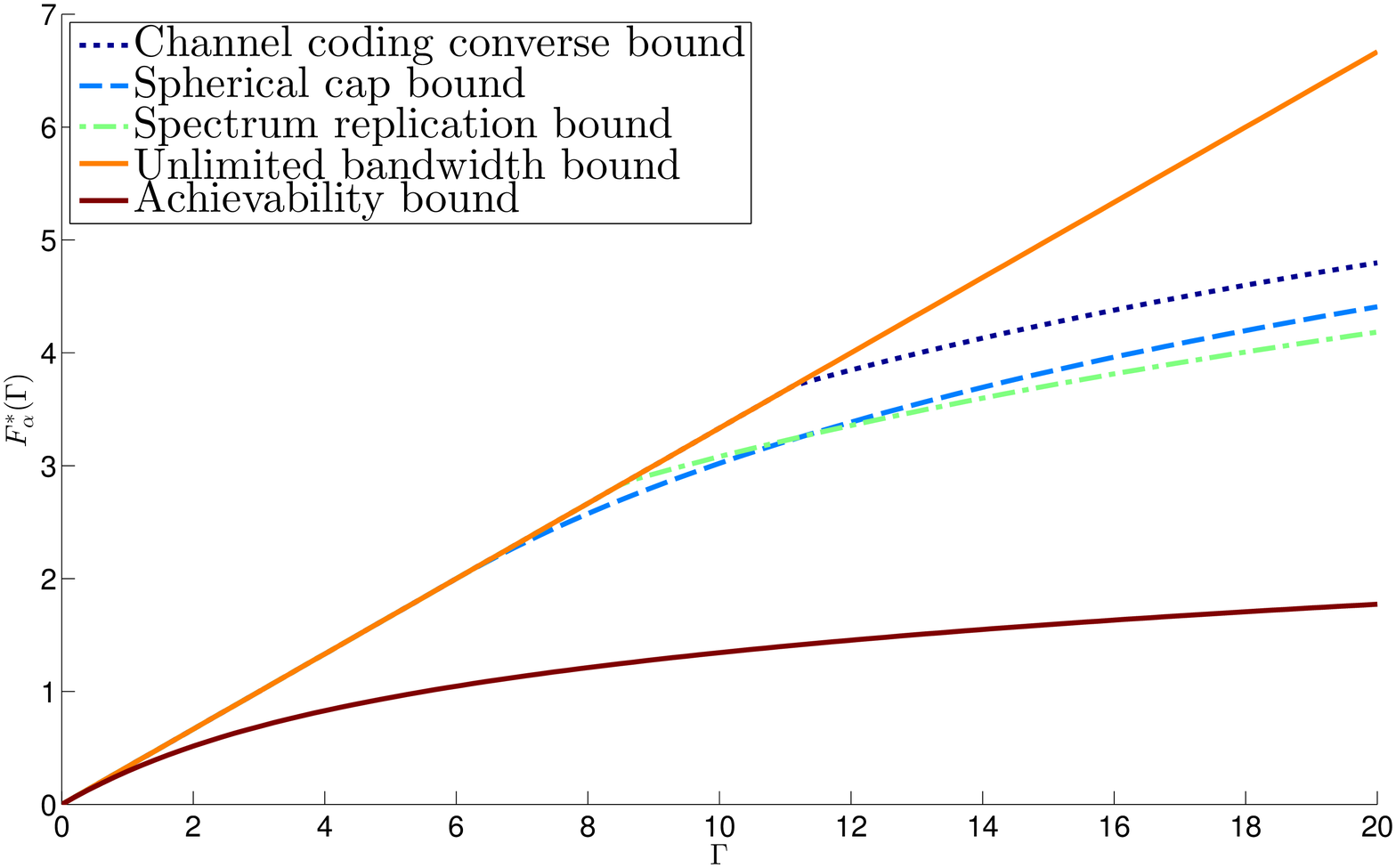}\protect\caption{Various bounds on $F_{\alpha}^{*}(\Gamma)$ for $\alpha=2$. \label{fig:alpha=00003D2}}
\end{figure}

\begin{figure}
\centering{}\includegraphics[scale=0.4]{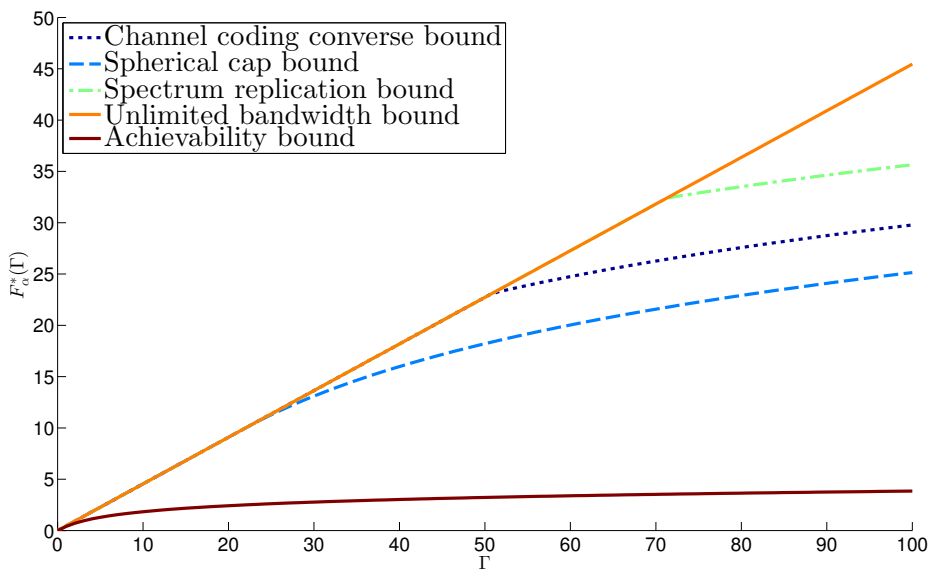}\protect\caption{Various bounds on $F_{\alpha}(\Gamma)$ for $\alpha=10$. \label{fig:alpha=00003D10}}
\end{figure}

In Figures \ref{fig:alpha=00003D0.1}-\ref{fig:alpha=00003D10}, the
values $\alpha=0.1,1,2,10$ are considered, and the channel coding
converse bound \eqref{eq: Channel Coding Upper Bound}, the spherical
cap bound \eqref{eq: Spherical Cap Bound}, and the spectrum replication
bound \eqref{eq: Superposition Bound} are plotted (using \eqref{eq: Burnashev bound}
to bound $\gamma_{\alpha}$). For the sake of comparison, the unlimited-bandwidth
converse bound \eqref{eq: Burnashev bound}, and the achievability
bound \eqref{eq: Channel Coding Lower Bound} are also plotted.

It is evident that for $\alpha=0.1$, the channel coding converse
bound dominates all other bounds; for $\alpha=1$ the spherical cap
bound is better for some values of $\Gamma$, but for most SNRs the
channel coding converse bound is the best; for $\alpha=2$ the spherical
cap bound is best for some values of $\Gamma$, but for most SNRs
the spectrum replication bound is the best; and, for $\alpha=10$
the spherical cap bound dominates all other bounds. 

\begin{figure}
\centering{}\includegraphics[scale=0.4]{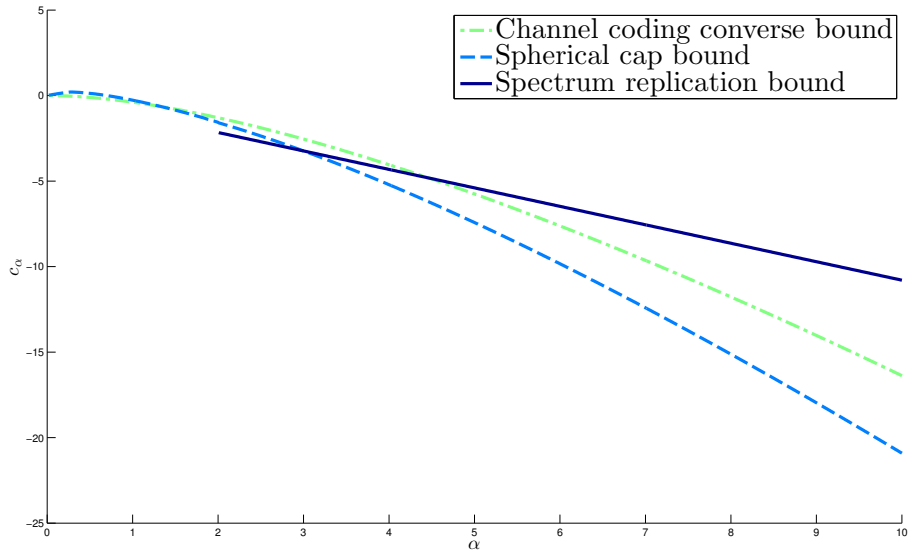}\protect\caption{$c_{\alpha}$ for the various bounds, when using \eqref{eq: Burnashev bound}
as an upper bound to $\gamma_{\alpha}$. \label{fig:High-SNR-constant }}
\end{figure}

To investigate systematically the behavior of the bounds for different
values of $\alpha$, we explore the high and low SNR regimes. At high
SNR, $\Gamma\to\infty$, it turns out that the all the converse bounds
have the same asymptotic form $\alpha\log(\Gamma)+c_{\alpha}+o(\Gamma)$,
for some $c_{\alpha}$. Thus, the various upper bounds differ by their
additive constant $c_{\alpha}$. The next proposition gives the value
of the constant $c_{\alpha}$. Its proof, as well as the proofs of
all the other propositions in this section can be found in Appendix
\ref{sec:SNR-Asymptotics }. 
\begin{prop}
\label{prop: High SNR constant}The converse bounds at $\Gamma\to\infty$
are given by
\begin{equation}
F_{\alpha}(\Gamma)\leq\alpha\log(\Gamma)+c_{\alpha}+o(\Gamma)\label{eq: High SNR bounds form}
\end{equation}
with 

\emph{
\begin{equation}
c_{\alpha}=\begin{cases}
\alpha-(1+\alpha)\log(1+\alpha), & \mbox{Channel coding converse bound (Prop. \ref{prop:Channel coding upper bound})}\\
\alpha\log\left(\frac{\gamma_{\alpha}}{\alpha}\right)+\alpha, & \mbox{Spherical cap bound (Thm. \ref{thm:Spherical cap bound})}\\
\alpha\log\left(\frac{e}{8}\right), & \mbox{Spectrum replication bound (Thm. \ref{thm: spectrum replication bound} ), }\alpha\geq2.
\end{cases}\label{eq: high SNR costant}
\end{equation}
}For $\alpha<2$ the spectrum replication bound of Theorem \ref{thm: spectrum replication bound}
increases linearly with $\Gamma$, and is thus useless for high SNR. 
\end{prop}
Fig. \ref{fig:High-SNR-constant } shows the value of $c_{\alpha}$
versus $\alpha$. As can be seen, for $0<\alpha\leq1.34$, the channel
coding converse bound has the best constant, for $1.34<\alpha\leq2$
and $\alpha\geq3$, the spherical cap bound has the best constant,
and for $2<\alpha\leq3$, the spectrum replication bound has the best
constant. Nonetheless, if the bound \eqref{eq: Burnashev bound} is
not really tight for $\alpha<2$, and its actual value is $\gamma_{\alpha}=\frac{\alpha}{2(1+\alpha)}$,
just as for $\alpha\geq2$, then the spectrum replication bound would
be the best for all $\alpha\leq3$ (see Remark \ref{rem:Conjecture based on asymptotic analysis}).
We remark that the DPT based bound \eqref{eq: DPT band-limited},
given by
\begin{equation}
F_{\alpha}(\Gamma)\leq\alpha\cdot\log(1+\Gamma),\label{eq: DPT bound 2}
\end{equation}
is not displayed in Figures \ref{fig:alpha=00003D0.1}-\ref{fig:alpha=00003D10},
since it is worse than the best all other converse bounds. For high
SNR, this is also evident from Fig. \ref{fig:High-SNR-constant },
by noting that $c_{\alpha}<0$ if we take the minimum over of all
bounds (cf. \eqref{eq: High SNR bounds form} and \eqref{eq: DPT bound 2}).
Regarding the achievability bound of Proposition \ref{prop:Channel coding lower bound},
a slightly weaker statement can be made. 
\begin{prop}
\label{prop:High SNR behviour channel coding lower bound}The achievability
bound of Prop. \ref{prop:Channel coding lower bound} scales as $[1+o(\Gamma)]\cdot\alpha\log(\Gamma)$
as $\Gamma\to\infty$. 
\end{prop}
At the other extreme, at low SNR ($\Gamma\to0$), it is apparent that
just like in channel coding, the bandwidth constraint is immaterial,
and the performance of band-limited systems approaches that of unlimited-bandwidth
systems. In this regime, the additional dimensions offered by a possible
increase of the bandwidth do not improve the exponent, because the
increase in the MP$\alpha$E exponent due to the additional dimensions
is lower than the decrease in the exponent due to energy reduction
in the original dimensions. Proposition \ref{prop:Low SNR behviour channel coding lower bound}
describes the behavior of the channel coding converse bound for small
$\Gamma$. 
\begin{prop}
\label{prop:Low SNR behviour channel coding lower bound}The channel
coding converse bound of Prop. \ref{prop:Channel coding lower bound}
scales as $\frac{\alpha}{2(1+\alpha)}\Gamma+\Theta(\Gamma^{2})$ as
$\Gamma\to0$.
\end{prop}
For $\alpha\geq2$ the channel coding converse bound is linear in
$\Gamma$, and has the same slope as the unlimited-bandwidth converse
bound $\gamma_{\alpha}\Gamma$ (see \eqref{eq: Burnashev bound}).
For $\alpha<2$, however, there is still a gap. 

\begin{figure}
\centering{}\includegraphics[scale=0.4]{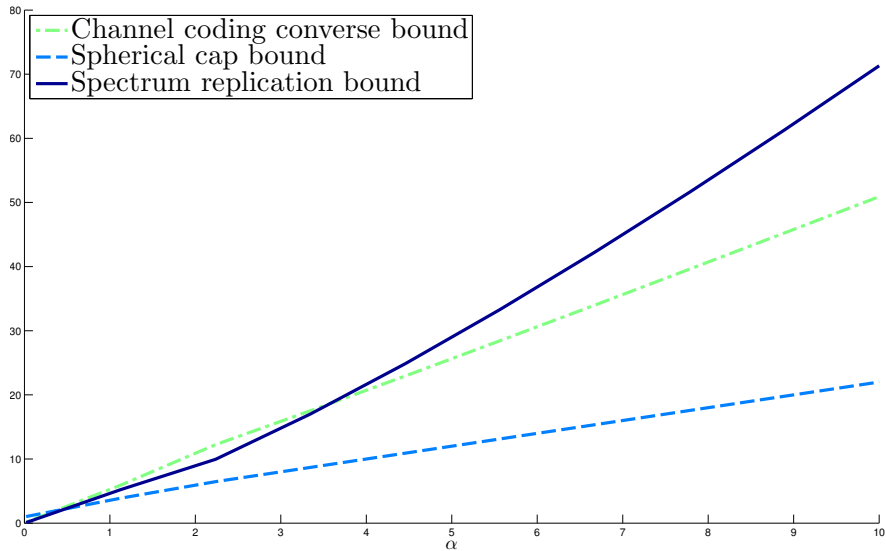}\protect\caption{The minimal SNR for which a band-limited bound deviates from the unlimited-bandwidth
bound, as a function of $\alpha$.\label{fig: Critical SNR }}
\end{figure}

Nevertheless, as the SNR increases, the band-limited exponent should
be strictly less than the unlimited-bandwidth exponent.\textbf{ }From
this aspect, an interesting figure merit for a bound is the minimal
SNR for which the bound deviates from\textbf{ }the unlimited-bandwidth
bound. For the spherical cap bound, this SNR is clearly $\Gamma_{\st[sc]}\dfn\frac{\alpha}{\gamma_{\alpha}}$.
For the channel coding converse bound, such an SNR $\Gamma_{\st[cc]}$
exists, but it is difficult to find it analytically. Indeed, as $\Gamma\to0$,
we get $\beta_{0}\to1$ and the channel coding converse bound reads\textbf{
\[
F_{\alpha}(\Gamma)\leq\frac{\alpha}{1+\alpha}\Gamma+\Theta(\Gamma^{2}),
\]
}and as evident from \eqref{eq: Burnashev bound}, the minimization
in \eqref{eq: Channel Coding Upper Bound} is dominated by the term
$\gamma_{\alpha}\Gamma$. For the spectrum replication bound, the
minimal SNR $\Gamma_{\st[sp]}$ for which the bound deviates from
the unlimited-bandwidth bound is also difficult to find analytically%
\footnote{The existence of such an SNR is also difficult to prove. Note that
$\Lambda_{\alpha}(\Gamma)=0$. Thus, if $\Lambda_{\alpha}(\Gamma)$
is a convex function of $\Gamma$ then a critical SNR $\Gamma_{\st[sp]}$,
such that $\Lambda_{\alpha}(\Gamma)>0$ for all $\Gamma>\Gamma_{\st[sp]}$
does exist. In turn, $\Lambda_{\alpha}(\Gamma)$ is the pointwise
supremum of $\frac{\Phi(\rho,\Gamma)-\gamma_{\alpha}\Gamma}{\rho}$,
and so if $\Phi(\rho,\Gamma)$ is a convex function of $\Gamma$,
then so is $\Lambda_{\alpha}(\Gamma)$. Unfortunately, verifying that
$\Phi(\rho,\Gamma)$ is a convex function of $\Gamma$ is not a trivial
task. Nonetheless, we were not able to find any counterexample for
the convexity of $\Phi(\rho,\Gamma)$. %
}. Thus, numerical results are displayed in Fig. \ref{fig: Critical SNR }.
From this aspect, it is seen that the spherical cap bound is usually
better than the two other bounds, except for very low values of $\alpha$. 

It is also interesting to note that for a given $\Gamma$, all bounds
tend to zero as $\alpha\to0$. For $\alpha\to\infty$ the spectrum
replication bound is useless, whereas the channel coding and spherical
cap bounds tend to $\frac{\Gamma}{2}$; the latter being the channel
capacity of the unlimited-bandwidth channel (per unit time per unit
bandwidth).

\appendices{\numberwithin{equation}{section}}

\section{Proofs of Converse Bounds\label{sec:Proof-of-Converse-Bounds}}
\begin{IEEEproof}[Proof of Theorem \ref{thm:Spherical cap bound}]
As said in Section \ref{sec:System model}, any band-limited signal
$s(t,u)$ of energy $PT$ can be identified with a vector $\mathbf{s}(u)\dfn(s_{1}(u),\ldots s_{N}(u))\in\mathbb{R}^{N}$,
where $N=2WT$ (see \eqref{eq: band limited}). Due to the power constraint,
$\mathbf{s}(u)$ lies on the surface of the of radius $\sqrt{PT}$,
centered at the origin. 

We begin with a few definitions. With some abuse of notation, the
system ${\cal S}_{N}$ will be identified with the projection vectors
of the signals in ${\cal S}_{T}$, i.e., ${\cal S}_{N}\dfn\left\{ \mathbf{s}(u):\; u\in[0,1)\right\} $,
and its MP$\alpha$E will be denoted by $e_{\alpha}({\cal S}_{N})$,
where the estimator will be understood from context. We denote the
set of parameters values pertaining to a signal subset $\overline{{\cal S}}_{N}\subseteq{\cal S}_{N}$,
by $\mathbb{U}(\overline{{\cal S}}_{N})$, i.e., $u\in\mathbb{U}(\overline{{\cal S}}_{N})$
iff $\mathbf{s}(u)\in\overline{{\cal S}}_{N}$. Also, we denote by
$\left|\mathbb{U}(\overline{{\cal S}}_{N})\right|$ the standard Lebesgue
measure of the set $\mathbb{U}(\overline{{\cal S}}_{N})\subseteq[0,1)$.
Furthermore, for any unit vector $\mathbf{g}\in\mathbb{R}^{N}$ and
an angle $\theta\in[0,\pi]$ we define the \emph{spherical cap} as
\[
{\cal A}_{\theta}(\mathbf{g},{\cal S}_{N})\dfn\left\{ \mathbf{s}(u)\in{\cal S}_{N}:\;\left\langle \mathbf{s}(u),\mathbf{g}\right\rangle \geq\sqrt{PT}\cdot\cos\theta\right\} ,
\]
where, as usual, the inner product is defined as $\left\langle \mathbf{s}(u),\mathbf{g}\right\rangle \dfn\sum_{n=1}^{N}s_{n}(u)\cdot g_{n}$.
We begin with the following \emph{measuring} argument.
\begin{lem}
\label{lem:spherical cap counting argument}Let $\theta\in(0,\frac{\pi}{2})$
and ${\cal S}_{N}$ be given. Then, there exists unit vector $\mathbf{g}$
for which 
\begin{equation}
\left|{\cal \mathbb{U}}\left[{\cal A}_{\theta}(\mathbf{g},{\cal S}_{N})\right]\right|\geq\exp\left\{ \frac{N}{2}\cdot\left[\log\left(\sin^{2}\theta\right)-\Theta\left(\frac{\log N}{N}\right)\right]\right\} .\label{eq: measuring argument statment}
\end{equation}
\end{lem}
\begin{IEEEproof}[Proof of Lemma \ref{lem:spherical cap counting argument}]
 The idea of the proof is similar to \cite[pp. 293-294]{wyner1973bound}.
Let $A_{N}(\theta)$ denote the surface area of a spherical cap of
angle $\theta$ on a sphere of radius $\sqrt{PT}$, in an $N$ dimensional
space.\textbf{ }Note that $A_{N}(\pi)$ is the surface area of the
entire sphere. Now, define 
\[
\Xi\dfn\int_{0}^{1}\int_{{\cal B}_{N}}\I\left\{ \left\langle \mathbf{s}(u),\mathbf{g}\right\rangle \geq S\cdot\cos\theta\right\} \d{\cal B}_{N}(\mathbf{g})\cdot\d u
\]
where ${\cal B}_{N}$ is the surface of the $N$-dimensional unit
sphere and $\d{\cal B}_{N}(\mathbf{g})$ is a differential surface
area around $\mathbf{g}$. On the one hand, $\Xi$ is trivially given
by $A_{N}(\theta)$. On the other hand, using Fubini's theorem \cite[Chapter 18]{billingsley2012probability},
$\Xi$ can also be expressed with the integration order exchanged,
and so 
\begin{align}
\Xi & =\int_{{\cal B}_{N}}\int_{0}^{1}\I\left\{ \left\langle \mathbf{s}(u),\mathbf{g}\right\rangle \geq S\cdot\cos\theta\right\} \d u\cdot\d{\cal B}_{N}(\mathbf{g})\\
 & =\int_{{\cal B}_{N}}\left|\mathbb{U}\left[{\cal A}_{\theta}(\mathbf{g},{\cal S}_{N})\right]\right|\d{\cal B}_{N}(\mathbf{g})\\
 & \leq A_{N}(\pi)\cdot\max_{\mathbf{g}\in{\cal B}_{N}}\left|{\cal \mathbb{U}}\left[{\cal A}_{\theta}(\mathbf{g},{\cal S}_{N})\right]\right|.
\end{align}
Thus, there exists $\mathbf{g}\in{\cal B}_{N}$ such that 
\[
\left|\mathbb{U}\left[{\cal A}_{\theta}(\mathbf{g},{\cal S}_{N})\right]\right|\geq\frac{A_{N}(\theta)}{A_{N}(\pi)}.
\]
To conclude, we use \cite[eqs. (27) and (28)]{shannon1959probability}
\[
\frac{A_{N}(\theta)}{A_{N}(\pi)}=\frac{\exp\left[N\log\left(\sin\theta\right)\right]}{\sqrt{2\pi N}\cdot\sin\theta\cdot\cos\theta}\cdot\left[1+O\left(\frac{1}{N}\right)\right]=\exp\left\{ \frac{N}{2}\left[\log\left(\sin^{2}\theta\right)-\frac{\log(N)}{N}-\frac{O(1)}{N}\right]\right\} .
\]

\end{IEEEproof}
Let ${\cal S}_{N}$ be given, and denote its estimator by $\hat{u}(\mathbf{y})$.
In addition, let $\mathbf{g}$ be a unit vector that satisfies \eqref{eq: measuring argument statment},
and let ${\cal U}_{N}\dfn\mathbb{U}\left[{\cal A}_{\theta}(\mathbf{g},{\cal S}_{N})\right]$
the corresponding parameter values of its spherical cap. We shall
now construct from ${\cal S}_{N}$, two modulation-estimation systems,
${\cal \overline{S}}_{N}$ and $\tilde{{\cal S}}_{N}$, using signals
only from ${\cal A}_{\theta}(\mathbf{g},{\cal S}_{N})$, such that
\begin{equation}
e_{\alpha}({\cal S}_{N})\geq e_{\alpha}(\overline{{\cal S}}_{N})\geq\left|{\cal U}_{N}\right|^{\alpha}\cdot e_{\alpha}(\tilde{{\cal S}}_{N}).\label{eq: relation between systems}
\end{equation}
Now, although $\tilde{{\cal S}}_{N}$ is band-limited just like ${\cal S}_{N}$,
we will bound $e_{\alpha}(\tilde{{\cal S}}_{N})$ using the \emph{unlimited-bandwidth
}bound. This and \eqref{eq: relation between systems} will provide
a bound on $F_{\alpha}(\Gamma)$. 

To construct $\overline{{\cal S}}_{N}$, we shall map ${\cal U}_{N}$
onto $\overline{{\cal U}}_{N}\dfn[0,|{\cal U}_{N}|)$ in an order
preserving manner (see Fig. \ref{fig: spherical cap proof}). For
example, if ${\cal U}_{N}=\bigcup_{i=1}^{I}{\cal I}_{i}$, where ${\cal I}_{i}$
are disjoint intervals of the form $[a_{i},b_{i})$, and $a_{1}<b_{1}\leq a_{2}<\cdots<a_{I}\leq b_{I}$
then such a mapping is easily obtained by eliminating the spaces between
every two consecutive intervals. Indeed, at the first step, the interval
${\cal I}_{I}$ will be shifted by $a_{I}-b_{I-1}$ to the left, such
that ${\cal I}_{I-1}$ and ${\cal I}_{I}$ are combined into a single
interval ${\cal I}_{I-1}^{(1)}$, while ${\cal I}_{i}^{(1)}={\cal I}_{i}$
is set for $1\leq i<I-1$. At the second step, the interval ${\cal I}_{I-1}^{(1)}$
is combined with ${\cal I}_{I-2}^{(1)}$ to a single interval ${\cal I}_{I-2}^{(2)}$
in the same manner. Continuing in this manner for $I-1$ steps, we
obtain a single interval, which can be translated to $[0,|{\cal U}_{N}|)$.
More generally, it is easy to verify that the mapping
\begin{equation}
\Psi[u]\dfn\int_{0}^{u}\I\left[w\in{\cal U}_{N}\right]\d w\label{eq: contraction mapping}
\end{equation}
satisfies the required properties. Note that the integral in \eqref{eq: contraction mapping}
exists since the mapping $u\to s(t,u)$ is assumed to be measurable.
The function $\Psi[\cdot]$ is monotonic and Lipschitz continuous
with constant $1$ as 
\[
\left|u_{1}-u_{2}\right|\geq\left|\Psi[u_{1}]-\Psi[u_{2}]\right|
\]
for any $u_{1},u_{2}\in[0,1)$. So, using the estimator $\overline{u}(\mathbf{y})\dfn\Psi[\hat{u}(\mathbf{y})]$
for $\overline{{\cal S}}_{N}$, we have 
\begin{equation}
\E_{u}\left\{ \left|\hat{u}(\mathbf{Y})-u\right|^{\alpha}\right\} \geq\E_{\Psi[u]}\left\{ \left|\Psi[\hat{u}(\mathbf{Y})]-\Psi[u]\right|^{\alpha}\right\} ,\label{eq: contraction bound}
\end{equation}
for any $u\in{\cal U}_{N}$, where in the left-hand side (right-hand
side) the system ${\cal S}_{N}$ (respectively, $\overline{{\cal S}}_{N}$)
is assumed. Hence, 
\begin{equation}
e_{\alpha}({\cal S}_{N})\geq e_{\alpha}(\overline{{\cal S}}_{N}).\label{eq: relation between first two systems}
\end{equation}
Now, consider the signal set
\begin{equation}
{\cal \tilde{S}}_{N}\dfn\left\{ \tilde{\mathbf{s}}(u)=\mathbf{s}(u)-\left\langle \mathbf{s}(u),\mathbf{g}\right\rangle \cdot\mathbf{g}:\;\mathbf{s}(u)\in{\cal A}_{\theta}(\mathbf{g},{\cal S}_{N})\right\} .\label{eq: projections set}
\end{equation}
To wit, geometrically, this is the signal set obtained by removing
the projecting of the signal vector $\mathbf{s}(u)$ onto $\mathbf{g}$
from $\mathbf{s}(u)$. Clearly, for any $\mathbf{s}(u)\in{\cal A}_{\theta}(\mathbf{g},{\cal S}_{N})$
and it corresponding $\tilde{\mathbf{s}}(u)\in\tilde{{\cal S}}_{N}$
according to \eqref{eq: projections set}, 
\begin{align}
PT & =\left\Vert \mathbf{s}(u)\right\Vert ^{2}\\
 & =\left\Vert \mathbf{s}(u)-\left\langle \mathbf{s}(u),\mathbf{g}\right\rangle \cdot\mathbf{g}+\left\langle \mathbf{s}(u),\mathbf{g}\right\rangle \cdot\mathbf{g}\right\Vert ^{2}\\
 & \trre[=,a]\left\Vert \tilde{\mathbf{s}}(u)\right\Vert ^{2}+\left\Vert \left\langle \mathbf{s}(u),\mathbf{g}\right\rangle \cdot\mathbf{g}\right\Vert ^{2}\\
 & \geq\left\Vert \tilde{\mathbf{s}}(u)\right\Vert ^{2}+PT\cdot\cos^{2}\theta,
\end{align}
where $(a)$ follows from the Pythagorean theorem and the orthogonality
of $\tilde{\mathbf{s}}(u)$ and $\mathbf{g}$. Thus, the signal set
$\tilde{{\cal S}}_{N}$ satisfies an energy constraint of $PT[1-\cos^{2}\theta]=PT\cdot\sin^{2}\theta$. 

We can now construct a modulation-estimation system which is based
on the signal set $\tilde{{\cal S}}_{N}$ and the original domain
of the parameter. This is simply done by scaling back $\overline{{\cal U}}_{N}=[0,|{\cal U}_{N}|)$
to the original interval $[0,1)=|{\cal U}_{N}|^{-1}\cdot[0,|{\cal U}_{N}|)$.
The system operates as follows. To modulate a parameter $v\in[0,1)$,
first $u(v)$ is set to 
\[
u(v)=\Psi^{-1}\left[|{\cal U}_{N}|^{-1}\cdot v\right],
\]
i.e. the parameter $v\in[0,1)$ is first mapped to $\overline{{\cal U}}_{N}$
and then mapped to the $u\in[0,1)$ that satisfies $\mathbf{s}\left(u(v)\right)\in{\cal A}_{\theta}(\mathbf{g},{\cal S}_{N})$.
Then, $s\left(t,u(v)\right)=\sum s_{n}\left(u(v)\right)\cdot\phi_{n}(t)$
is transmitted over the channel \eqref{eq: Gaussian channel}. The
estimator $\hat{v}(\mathbf{y})$ of $v$, is given by
\[
\hat{v}(\mathbf{y})=|{\cal U}_{N}|^{-1}\cdot\overline{u}(\mathbf{y}).
\]
Now, due to the scaling operation from $\overline{{\cal S}}_{N}$
to $\tilde{{\cal S}}_{N}$ by a factor of $|{\cal U}_{N}|^{-1}$,
the ratio between their MP$\alpha$E's is not larger than $|{\cal U}_{N}|^{-\alpha}$,
to wit, for any given $v\in[0,1)$
\begin{equation}
\E_{v}\left\{ \left|\hat{v}(\mathbf{Y})-v\right|^{\alpha}\right\} \leq|{\cal U}_{N}|^{-\alpha}\cdot\E_{|{\cal U}_{N}|\cdot v}\left\{ \left|\overline{u}(\mathbf{y})-|{\cal U}_{N}|\cdot v\right|^{\alpha}\right\} \label{eq: streching bound}
\end{equation}
where in the left-hand side (right-hand side) the system $\tilde{{\cal S}}_{N}$
(respectively, $\overline{{\cal S}}_{N}$) is assumed. This, together
with \eqref{eq: relation between first two systems} implies \eqref{eq: relation between systems}. 

Now, we note that the modulation-estimation system for $v$ has a
power limitation of $PT\cdot\sin^{2}\theta$. So, a lower bound for
the MP$\alpha$E of \emph{unlimited-bandwidth} systems can be used
to obtain to lower bound the left-hand side of \eqref{eq: streching bound},
and hence, 
\[
\E_{v}\left\{ \left|\hat{v}(\mathbf{Y})-v\right|^{\alpha}\right\} \geq\exp\left[-\frac{N}{2}\gamma_{\alpha}\Gamma\sin^{2}\theta\right].
\]
This, along with \eqref{eq: relation between systems} and Lemma \ref{lem:spherical cap counting argument},
implies that for any given $\delta>0$, there exists $N$ sufficiently
large such that
\[
\E_{u}\left\{ \left|\hat{u}(\mathbf{Y})-u\right|^{\alpha}\right\} \geq\exp\left\{ \frac{N}{2}\cdot\alpha\left[\log\left(\sin^{2}\theta\right)-\delta\right]\right\} \exp\left[-\frac{N}{2}\gamma_{\alpha}\Gamma\sin^{2}\theta\right].
\]
Now, the angle $\theta\in(0,\frac{\pi}{2})$ is arbitrary, and thus
can be optimized. Denoting $\tau\dfn\sin^{2}\theta$ we get
\[
\E_{u}\left\{ \left|\hat{u}(\mathbf{Y})-u\right|^{\alpha}\right\} \geq\exp\left\{ -\frac{N}{2}\left[\sup_{0<\tau<1}\left(\gamma_{\alpha}\Gamma\tau-\alpha\log\tau\right)-\alpha\delta\right]\right\} ,
\]
and after maximizing over $\tau$, and taking $\delta\downarrow0$,
\eqref{eq: Spherical Cap Bound} is immediately obtained. 
\end{IEEEproof}

\begin{IEEEproof}[Proof of Theorem \ref{thm: spectrum replication bound}]
 Let $\delta>0$ be given. As in the proof of the spherical cap bound,
let a signal set ${\cal S}_{N}\dfn\left\{ \mathbf{s}(u):\; u\in[0,1)\right\} $
be given. As was discussed in Section \ref{sec:System model}, for
\emph{any} given dimension, we can transform a vector $\mathbf{s}(u)$
to a signal $s(t,u)$ using an orthonormal basis $\{\phi_{n}(t)\}$.
Specifically, let us consider an orthonormal basis of $L\dfn MN$
signals $\{\phi_{l}(t),\;0\leq t\leq T\}_{l=1}^{L}$, where $M\gg1$,
and $M$ integer. We assume that the system ${\cal S}_{N}$ uses $\{\phi_{l}(t),\;0\leq t\leq T\}_{l=1}^{N}$
to transform $\mathbf{s}(u)\in\mathbb{R}^{N}$ to a signal $s(t,u)$.
We now construct a new system, that modulates a parameter $v\in[0,1)$,
using a signal set $\tilde{{\cal S}}_{N}\in\mathbb{R}^{L}$ , which
is transformed to a signal using $\{\phi_{l}(t),\;0\leq t\leq T\}_{l=1}^{L}$.
Since $N=2WT$ and $L=2MWT$, as said in\textbf{ }Section\textbf{
}\ref{sec:Converse-Bounds},\textbf{ }one can think of the system
$\tilde{{\cal S}}_{N}$ as using bandwidth $MW\gg W$. Its total frequency
band $[0,MW)$ is partitioned into $M$ consecutive frequency bands
$[0,W)$, $[W,2W)$,..., $[(M-1)W,MW)$, and the value of the parameter
is modulated using both the choice of active frequency band $0\leq m\leq M-1$%
\footnote{Nothing it transmitted at all other bands. However, as discussed in
Section \ref{sec:System model} the system ${\cal S}_{T}$ is, in
essence, only approximately band-limited to $[0,W)$, and thus its
signals have out-of-band energy. In the frequency position modulation
described here, this could create interference between neighboring
frequency bands. However, since we eventually bound the MP$\alpha$E
of $\tilde{{\cal S}}_{T}$ by a bound for an unlimited-bandwidth system,
our proof remains in tact even if we choose the modulation frequencies
to be $f(m)=mKW$ , for any arbitrarily large integer $K$. Hence,
the effect of the interference can made negligible.%
}, and the specific signal within the band. 

We now describe the system $\tilde{{\cal S}}_{N}$ which modulates
$v\in[0,1)$. Let $v_{\st[c]}=\frac{\left\lfloor M\cdot v\right\rfloor }{M}$,
where $\left\lfloor \cdot\right\rfloor $ is the floor operation,
and $v_{\st[r]}=v-v_{\st[c]}$. The idea is to use $v_{\st[c]}$ to
choose one of $M$ possible sets of basis functions $\left\{ \{\phi_{l}\}_{l=1}^{N},\{\phi_{l}\}_{l=N+1}^{2N},\ldots\{\phi_{l}\}_{l=(M-1)N+1}^{MN}\right\} $,
and to use $v_{\st[r]}$ to choose which vector to transmit over the
chosen $N$ basis functions, while utilizing the original system ${\cal S}_{N}$.

The modulation-estimation system $\tilde{{\cal S}}_{N}$ is depicted
in Fig. \ref{fig:Modulation-Estimation-system super position} (in
continuous time). 
\begin{figure}
\begin{centering}
\includegraphics[scale=0.8]{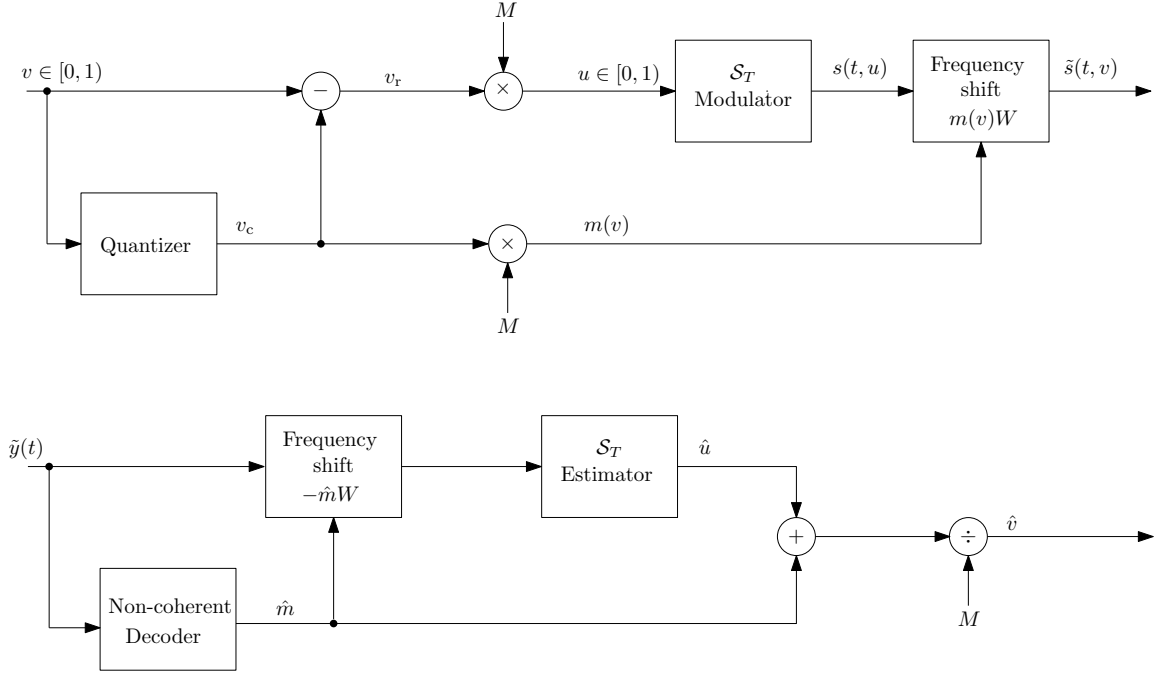}
\par\end{centering}

\protect\caption{Modulation-Estimation system for the proof of the spectrum replication
bound.\label{fig:Modulation-Estimation-system super position}}

\end{figure}
 Specifically, to modulate $v$, a modulation index is chosen using
the coarse part as
\[
m(v)\dfn M\cdot v_{\st[c]}\in\{0,\ldots,M-1\},
\]
and a vector of coefficients is chosen as $\mathbf{s}(M\cdot v_{\st[r]})$.
Then, $\tilde{\mathbf{s}}(v)$, the coefficient vector of $\tilde{{\cal S}}_{N}$,
is chosen with the entries
\[
\tilde{s}_{l}(v)=\begin{cases}
s_{l-m(v)}(M\cdot v_{\st[r]}), & m(v)N+1\leq l\leq m(v)N+N\\
0, & \mbox{otherwise}
\end{cases}
\]
(note that $M\cdot v_{\st[r]}\in[0,1)$), and $\tilde{s}(t,v)=\sum_{l=1}^{L}\tilde{s}_{l}(v)\cdot\phi_{l}(t)$
is transmitted over the channel \eqref{eq: Gaussian channel}. To
wit, only the signals $\{\phi_{l}(t)\}_{l=m(v)N+1}^{m(v)N+N}$, which
represent the frequency band $[m(v)W,(m(v)+1)W)$, have non-zero coefficients. 

At the receiver, a proper projection vector $\tilde{\mathbf{y}}$
is obtained as in \eqref{eq: output signal projection}, but this
time, over $L$ basis functions. Specifically, we define the projections
\[
\tilde{y}_{m,n}\dfn\int_{0}^{T}y(t)\cdot\tilde{\phi}_{mN+n}(t)\cdot\d t\quad0\leq m\leq M-1,\quad1\leq n\leq N,
\]
and $\tilde{\mathbf{y}}_{m}\dfn(\tilde{y}_{m,1},\ldots,\tilde{y}_{m,N})$,
as well as the (scaled) energies $q_{m}\dfn\frac{2}{N_{0}}\left\Vert \tilde{\mathbf{y}}_{m}\right\Vert ^{2}=\frac{2}{N_{0}}\sum_{n=1}^{N}\tilde{y}_{m,n}^{2}.$
The estimator $\hat{v}(\tilde{\mathbf{y}})$ of $\tilde{{\cal S}}_{N}$
is obtained in two steps. In the first, we decode $m(v)$, using a
non-coherent decoder, which decides based on the \emph{maximum projection
energy}, i.e. 
\begin{equation}
\hat{m}(\tilde{\mathbf{y}})\dfn\argmax_{m\in\{0,\ldots,M-1\}}q_{m}.\label{eq: non coherent decoder}
\end{equation}
In the second step, we estimate the parameter $v$ as
\[
\hat{v}(\tilde{\mathbf{y}})=\frac{\hat{m}+\hat{u}(\tilde{\mathbf{y}}_{\hat{m}})}{M},
\]
where $\hat{u}(\mathbf{y})$ is the estimator of the original system
${\cal S}_{N}$, and for clarity, the dependence of $\hat{m}$ on
$\tilde{\mathbf{y}}$ was omitted. In words, in the second step, we
assume that $\hat{m}$ is the correct index, and use the vector $\tilde{\mathbf{y}}_{\hat{m}}$
as the input to the estimator of ${\cal S}_{N}$. 

The exponential behavior of the MP$\alpha$E of the system $\tilde{{\cal S}}_{N}$
will be different from that of ${\cal S}_{N}$ only if $M$ increases
exponentially with $N$. Hence, we assume that $M\dfn\exp(TWR)=\exp(\frac{N}{2}\cdot R)$
for some `rate' $R>0$. In Appendix \ref{sec:Reliability-Analysis-of},
we analyze the reliability of the non-coherent decoder, using large
deviations analysis of chi-square random variables. Denoting the error
event, in the first step of the estimation, by ${\cal E}$, it is
shown there that for all $v\in[0,1)$

\begin{equation}
\P_{v}[{\cal E}]\leq\exp\left[-\frac{N}{2}\cdot G(\Gamma,R)\right]\label{eq: error probability decays with G}
\end{equation}
where
\[
G(\Gamma,R)\dfn\max_{0\leq\rho\leq1}\left\{ \Phi(\rho,\Gamma)-\rho R\right\} 
\]
and $\Phi(\rho,\Gamma)$ is as defined in \eqref{eq: Phi definition}.
The MP$\alpha$E of $\tilde{{\cal S}}_{N}$ is then bounded as follows.
For all $N$ sufficiently large
\begin{align}
\E_{v}\left\{ \left|\hat{v}(\mathbf{\tilde{Y}})-v\right|^{\alpha}\right\}  & =\P_{v}[{\cal E}]\cdot\E_{v}\left\{ \left|\hat{v}(\mathbf{\tilde{Y}})-v\right|^{\alpha}|{\cal E}\right\} +\P_{v}[{\cal E}^{c}]\cdot\E_{v}\left\{ \left|\hat{v}(\mathbf{\tilde{Y}})-v\right|^{\alpha}|{\cal E}^{c}\right\} \\
 & \leq\P_{v}[{\cal E}]+\E_{v}\left\{ \left|\hat{v}(\mathbf{\tilde{Y}})-v\right|^{\alpha}|{\cal E}^{c}\right\} \\
 & \leq\exp\left[-\frac{N}{2}\cdot G(\Gamma,R)\right]+\E_{v}\left\{ \left|\hat{v}(\mathbf{\tilde{Y}})-v\right|^{\alpha}|{\cal E}^{c}\right\} \\
 & =\exp\left[-\frac{N}{2}\cdot G(\Gamma,R)\right]+\E_{v}\left\{ \left.\left|\frac{\hat{m}(\tilde{\mathbf{Y}})}{M}+\frac{\hat{u}(\tilde{\mathbf{Y}}_{\hat{m}(\tilde{\mathbf{Y}})})}{M}-v_{\st[c]}-v_{\st[r]}\right|^{\alpha}\right|{\cal E}^{c}\right\} \\
 & \trre[\leq,a]\exp\left[-\frac{N}{2}\cdot G(\Gamma,R)\right]+2\cdot\E_{v}\left\{ \left|\frac{\hat{u}(\tilde{\mathbf{Y}}_{m(v)})}{M}-v_{\st[r]}\right|^{\alpha}\right\} \\
 & =\exp\left[-\frac{N}{2}\cdot G(\Gamma,R)\right]+\frac{2}{M^{\alpha}}\E_{u}\left\{ \left|\hat{u}(\tilde{\mathbf{Y}}_{m(v)})-M\cdot u_{\st[r]}\right|^{\alpha}\right\} \\
 & \leq\exp\left[-\frac{N}{2}\cdot G(\Gamma,R)\right]+2\cdot\exp\left\{ -\frac{N}{2}\cdot\left[\alpha R+F_{\alpha}(\Gamma)\right]\right\} \\
 & \leq2\cdot\exp\left\{ -\frac{N}{2}\cdot\min\left[G(\Gamma,R),\alpha R+F_{\alpha}(\Gamma)\right]\right\} .\label{eq: MSE exponents balance}
\end{align}
In $(a)$, we have used the fact that conditioned on ${\cal E}^{c}$,
we have $\frac{\hat{u}(\tilde{\mathbf{Y}}_{\hat{m}})}{M}=v_{\st[c]}$
and the fact that for $G(\Gamma,R)>0$ (which is our regime of interest),
\eqref{eq: error probability decays with G} implies that $\P_{v}[{\cal E}^{c}]\to1$
as $N\to\infty$. So, for any random variable $X$, and all $N$ sufficiently
large, 
\begin{align}
\E_{v}\left\{ X|{\cal E}^{c}\right\}  & =\frac{1}{\P_{v}[{\cal E}^{c}]}\E_{v}\left\{ X\cdot\I[X\in{\cal E}^{c}]\right\} \\
 & \leq\frac{1}{\P_{v}[{\cal E}^{c}]}\E_{v}[X]\\
 & \leq2\cdot\E_{v}[X].
\end{align}
Clearly, for any given $R\geq0$, the proposed system cannot achieve
an MP$\alpha$E exponent better than the converse bound of the unlimited-bandwidth
system, $\exp(-\frac{N}{2}\cdot\gamma_{\alpha}\Gamma)$. Hence, 
\begin{equation}
\min\left\{ G(\Gamma,R),\alpha R+F_{\alpha}(\Gamma)\right\} \leq\gamma_{\alpha}\Gamma,\label{eq: min of two exponent  upper bound}
\end{equation}
or, equivalently
\[
F_{\alpha}(\Gamma)\leq\begin{cases}
\gamma_{\alpha}\Gamma-\alpha R, & G(\Gamma,R)>\gamma_{\alpha}\Gamma\\
\infty, & G(\Gamma,R)\leq\gamma_{\alpha}\Gamma
\end{cases}.
\]
The relation $G(\Gamma,R)>\gamma_{\alpha}\Gamma$ can be easily seen
to be equivalent to $R<\Lambda_{\alpha}(\Gamma)$, where $\Lambda_{\alpha}(\Gamma)$
is defined in \eqref{eq: Lambda defintion}, and thus 
\[
F_{\alpha}(\Gamma)\leq\begin{cases}
\gamma_{\alpha}\Gamma-\alpha R, & R\leq\Lambda_{\alpha}(\Gamma)\\
\infty, & R>\Lambda_{\alpha}(\Gamma)
\end{cases}.
\]
Since $R\geq0$ is arbitrary, the tightest bound is obtained by choosing
$R=\Lambda_{\alpha}(\Gamma)$ which leads to \eqref{eq: Superposition Bound}.
\end{IEEEproof}

\section{Reliability Analysis of the Modulation Scheme of the Spectrum Replication
Bound\label{sec:Reliability-Analysis-of}}

\begin{figure}
\begin{centering}
\includegraphics[scale=0.7]{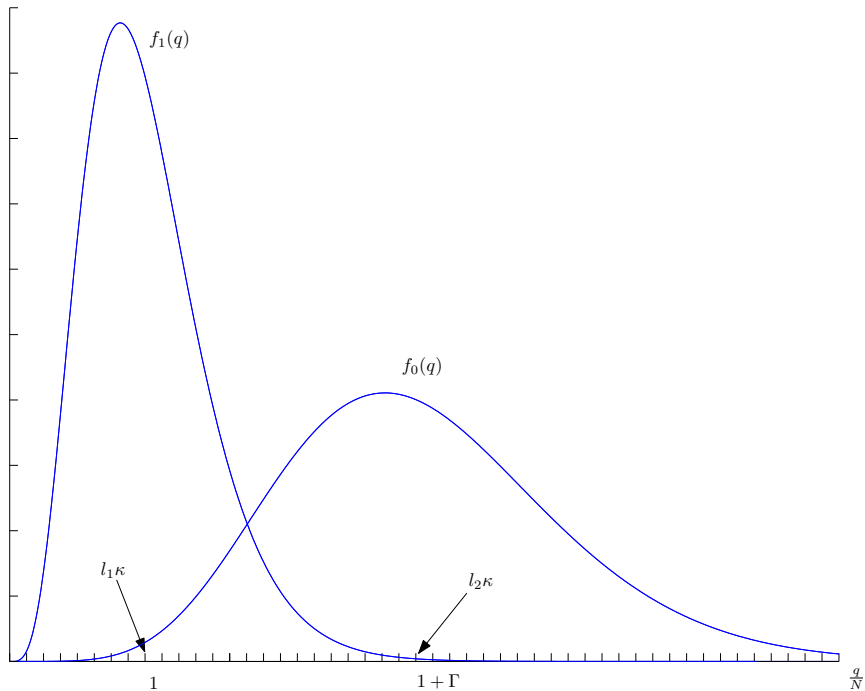}
\par\end{centering}

\protect\caption{The probability distribution functions of $Q_{0}$ and $Q_{1}$, for
$N=10$ and $\Gamma=2$.\label{fig:chi square}}
\end{figure}

In this appendix, we evaluate the reliability of the non-coherent
decoder \eqref{eq: non coherent decoder}. Let us denote the random
variables of the system by uppercase letters, e.g. $Q_{m}$. Due to
symmetry, it can be assumed w.l.o.g. that $m_{v}=0$. Then, it is
straightforward to verify that for any given $m\neq0$, we have $\tilde{Y}_{m,n}\sim{\cal N}(0,1)$
and $\{\tilde{Y}_{m,n}\}$ are independent. Consequently, $Q_{m}$
is a chi-square random variable of $N$ degrees of freedom. Similarly,
for $m=0$, we have $\tilde{Y}_{0,n}\sim{\cal N}(\varsigma_{M\cdot v_{\st[r]},n},1)$
i.e., $Q_{0}$ is a non-central chi-square random variable of $N$
degrees of freedom, and a non-centrality parameter $\lambda\dfn\frac{1}{\nicefrac{N_{0}}{2}}\sum_{n=1}^{N}\varsigma_{M\cdot v_{\st[r]},n}^{2}=\frac{2PT}{N_{0}}$.
We build on the analysis in \cite[Section 2.5, Section 2.12.2, Problem 2.14 and Problem 2.15]{viterbi2009principles}.
Let $f_{0}(\cdot)$ be the probability density function of $Q_{0}$
given that $m_{v}=0$. Then, for any $0\leq\rho\leq1$, the decoding
error probability can be bounded as
\begin{align}
\P_{v}[{\cal E}] & =1-\P\left[Q_{0}>Q_{m},\;\forall m\neq0|m_{v}=0\right]\\
 & =\int_{0}^{\infty}f_{0}(q)\left\{ 1-\left(\P[Q_{1}\leq q|m_{v}=0]\right)^{M-1}\right\} \cdot\d q\\
 & \leq M^{\rho}\int_{0}^{\infty}f_{0}(q)\P^{\rho}[Q_{1}>q|m_{v}=0]\cdot\d q\label{eq: error probability first step}
\end{align}
where the inequality is obtained from $1-(1-\alpha)^{M}\leq(M\alpha)^{\rho}$
(known as Gallager's union bound \cite[Lemma, p. 136]{gallager1968information}).
Let $\{K_{N}\}_{N=1}^{\infty}$ and $\{\kappa_{N}\}_{N=1}^{\infty}$
be two positive sequences which satisfy $K_{N}\uparrow\infty$ and
$\kappa_{N}\downarrow0$ as $N\to\infty$. The appropriate choices
for them will be discussed later on. For notational simplicity, let
us assume that $\nicefrac{K_{N}}{\kappa_{N}}$ is integer, and temporarily
omit the subscript $N$ in their notation. Then,
\begin{align}
\P_{v}[{\cal E}] & \leq M^{\rho}\int_{0}^{\infty}f_{0}(q)\cdot\P^{\rho}[Q_{1}>q|m_{v}=0]\cdot\d q\\
 & \leq M^{\rho}\sum_{l=0}^{\frac{K}{\kappa}-1}\int_{l\kappa N}^{(l+1)\kappa N}f_{0}(q)\cdot\P^{\rho}[Q_{1}>q|m_{v}=0]\cdot\d q+M^{\rho}\P\left(Q_{0}>KN|m_{v}=0\right)\\
 & \leq M^{\rho}\sum_{l=0}^{\frac{K}{\kappa}-1}\int_{l\kappa N}^{(l+1)\kappa N}f_{0}(q)\cdot\P^{\rho}[Q_{1}>l\kappa N|m_{v}=0]\cdot\d q+M^{\rho}\P\left(Q_{0}>KN|m_{v}=0\right)\\
 & =M^{\rho}\sum_{l=0}^{\frac{K}{\kappa}-1}\P^{\rho}[Q_{1}>l\kappa N|m_{v}=0]\cdot\P\left[l\kappa N\leq Q_{0}\leq(l+1)\kappa N|m_{v}=0\right]\nonumber \\
 & \hphantom{=}+M^{\rho}\P\left[Q_{0}>KN|m_{v}=0\right]\\
 & \leq M^{\rho}\frac{K}{\kappa}\cdot\left\{ \max_{0\le l\leq\frac{K}{\kappa}-1}\P^{\rho}[Q_{1}>l\kappa N|m_{v}=0]\cdot\P\left[l\kappa N\leq Q_{0}\leq(l+1)\kappa N|m_{v}=0\right]\right.\nonumber \\
 & \hphantom{=}\left.\vphantom{\max_{0\le l\leq\frac{K}{\kappa}-1}}+\P\left[Q_{0}>KN|m_{v}=0\right]\right\} ,\label{eq: error probability second step}
\end{align}
where in the last inequality, we have used the assumption that $\frac{K}{\kappa}>1$
for sufficiently large $N$. In order to evaluate the exponential
behavior of $\P_{v}[{\cal E}]$, it will be convenient to partition
the maximization to a few intervals. In each interval, we upper bound
the objective
\begin{equation}
\P^{\rho}[Q_{1}>l\kappa N|m_{v}=0]\cdot\P\left[l\kappa N\leq Q_{0}\leq(l+1)\kappa N|m_{v}=0\right]\label{eq: maximization argument}
\end{equation}
by an asymptotically tight upper bound. In essence, we are replacing
the probability of an interval by the tail probability of one of its
endpoints, according to the relative position of $l\kappa N$ w.r.t.
$\E[Q_{0}]=N$ and $\E[Q_{1}]=N+\lambda=N+\frac{2PT}{N_{0}}=N(1+\Gamma)$,
see Fig. \ref{fig:chi square}. 
Let $l_{1}$ be such that $l_{1}\kappa\leq1\leq(l_{1}+1)\kappa$ and
$l_{2}$ be such that $l_{2}\kappa\leq1+\Gamma\leq(l_{2}+1)\kappa$.
Then, for $l\leq l_{1}-1$ we upper bound \eqref{eq: maximization argument}
by
\begin{equation}
\P\left[Q_{0}\leq(l+1)\kappa N|m_{v}=0\right],\label{eq: tail bound 1}
\end{equation}
for $l_{1}\leq l\leq l_{2}-1$ we upper bound it by 
\begin{equation}
\P^{\rho}[Q_{1}>l\kappa N|m_{v}=0]\cdot\P\left[Q_{0}\leq(l+1)\kappa N|m_{v}=0\right],\label{eq: tail bound 2}
\end{equation}
for $l=l_{2}$ we upper bound it by
\begin{equation}
\P^{\rho}[Q_{1}>l\kappa N|m_{v}=0],\label{eq: tail bound 3}
\end{equation}
and for $l_{2}+1\leq l\leq\frac{K}{\kappa}-1$ we upper bound it by
\begin{equation}
\P^{\rho}[Q_{1}>l\kappa N|m_{v}=0]\cdot\P\left[Q_{0}\geq l\kappa N|m_{v}=0\right].\label{eq: tail bound 4}
\end{equation}
We can now analyze the behavior of the probabilities above, as $N\to\infty$.
To this end, we use the Chernoff bound for chi-square random variables,
using the known expressions for their moment generating functions
\cite[Section 19.8, eq. (19.45)]{lapidoth2009foundation}. For the
energy $Q_{1}$, we have that for $\eta\geq1$
\begin{align}
\P[Q_{1}\geq\eta N|m_{v}=0] & \leq\inf_{s\geq0}\frac{\E[e^{sQ_{1}}]}{e^{s\eta N}}\\
 & =\inf_{0\leq s<\nicefrac{1}{2}}\frac{\left(1-2s\right)^{-\nicefrac{N}{2}}}{e^{s\eta N}}\\
 & =\eta^{\nicefrac{N}{2}}\cdot\exp\left[-N\cdot\frac{\left(1-\frac{1}{\eta}\right)}{2}\right]\\
 & =\exp\left\{ -N\cdot\left[\frac{\eta-1-\log(\eta)}{2}\right]\right\} ,\label{eq: Q1 expnent}
\end{align}
where the critical point is $s=\frac{1}{2\eta}-\frac{1}{2}$, and
for $0<\eta<1$ we use the trivial bound
\[
\P[Q_{1}\geq\eta N|m_{v}=0]\leq1.
\]
In the same manner, for $Q_{0}$ and $0\leq\eta\leq1+\Gamma$ such
that $\E[Q_{1}]=N\leq\eta N\leq\E[Q_{0}]=N(1+\Gamma)$, 
\begin{align}
\P[Q_{0}\leq\eta N|m_{v}=0] & \leq\inf_{s\leq0}\frac{\E[e^{sQ_{0}}]}{e^{s\eta N}}\\
 & =\inf_{s\leq0}\frac{(1-2s)^{-\nicefrac{N}{2}}\cdot\exp\left[\frac{\lambda s}{1-2s}\right]}{e^{s\eta N}}\\
 & =\exp\left\{ -N\cdot\sup_{s\leq0}\left[-\frac{s}{1-2s}\Gamma+s\eta+\frac{\log(1-2s)}{2}\right]\right\} .
\end{align}
The critical point is 
\begin{align*}
\overline{s}(\eta) & \dfn\frac{1}{2}-\frac{1}{4\eta}-\sqrt{\left(\frac{1}{2}-\frac{1}{4\eta}\right)^{2}+\frac{1+\Gamma-\eta}{4\eta}}.
\end{align*}
After inserting back $\overline{s}(\eta)$, and straightforward algebra,
we get
\[
\P[Q_{0}\leq\eta N|m_{v}=0]\leq\exp\left\{ -N\cdot\frac{1}{2}\left(\eta+\Gamma+\log\left[\frac{\sqrt{4\eta\Gamma+1}+1}{2\eta}\right]-\sqrt{4\eta\Gamma+1}\right)\right\} .
\]
A similar analysis for $\eta>1+\Gamma$ gives a similar result, and
so 
\begin{equation}
\P[Q_{0}>\eta N|m_{v}=0]\leq\exp\left\{ -N\cdot\frac{1}{2}\left(\eta+\Gamma+\log\left[\frac{\sqrt{4\eta\Gamma+1}+1}{2\eta}\right]-\sqrt{4\eta\Gamma+1}\right)\right\} .\label{eq: Qo exponent}
\end{equation}
Returning to the error probability evaluation \eqref{eq: error probability second step}
and using the derived Chernoff bounds \eqref{eq: Q1 expnent} and
\eqref{eq: Qo exponent} in the bounds \eqref{eq: tail bound 1},
\eqref{eq: tail bound 2}, \eqref{eq: tail bound 3} and \eqref{eq: tail bound 4},
along with the continuity of the exponents in \eqref{eq: Q1 expnent}
and \eqref{eq: Qo exponent} %
\footnote{Note also that $\overline{s}(\eta)=0$ for $\eta=1+\Gamma$, which
leads to a trivial Chernoff bound (zero exponent).%
} we get
\begin{align}
G(\Gamma,R) & \geq-\frac{1}{\nicefrac{N}{2}}\log\P_{v}[{\cal E}]\\
 & \geq\min\left\{ \min_{0\leq\eta\leq1}\left\{ 2\cdot\left[-\frac{\overline{s}(\eta)}{1-2\overline{s}(\eta)}\Gamma+\overline{s}(\eta)\cdot\eta+\frac{\log(1-2\overline{s}(\eta))}{2}\right]+o(\kappa_{N})\right\} ,\right.\nonumber \\
 & \hphantom{\geq}\left.\min_{1\leq\eta\leq K_{N}}\left\{ 2\cdot\left[\frac{\rho[\eta-1-\log(\eta)]}{2}-\frac{\overline{s}(\eta)}{1-2\overline{s}(\eta)}\Gamma+\overline{s}(\eta)\cdot\eta+\frac{\log(1-2\overline{s}(\eta))}{2}\right]+o(\kappa_{N})\right\} ,\right.\nonumber \\
 & \hphantom{\geq}\left.2\cdot\left[-\frac{\overline{s}(K_{N})}{1-2\overline{s}(K_{N})}\Gamma+\overline{s}(K_{N})\cdot K_{N}+\frac{\log(1-2\overline{s}(K_{N}))}{2}\right]\right\} \nonumber \\
 & \hphantom{\geq}-\frac{2}{N}\log\frac{K_{N}}{\kappa_{N}}-\rho R.
\end{align}
Now, in the inner minimization, the second term decreases as $K_{N}$
increases, and the third term increases. Thus, for $N$ sufficiently
large, the third term will not be the minimal term. Also, since $\log\eta\leq1-\eta$
the first term is always not smaller than the second term. Hence,
the second term dominates the minimization. Choosing $K_{N}$ and
$\kappa_{N}$ such that $\lim_{N\to\infty}\frac{1}{N}\log\frac{K_{N}}{\kappa_{N}}=0$,
and optimizing over $0\leq\rho\leq1$ we obtain 
\[
G(\Gamma,R)\geq\max_{0\leq\rho\leq1}\left\{ \Phi(\rho,\Gamma)-\rho R\right\} 
\]
where $\Phi(\rho,\Gamma)$ is as defined in \eqref{eq: Phi definition}.
It is easy to verify that the objective function, in the optimization
problem pertaining to $\Phi(\rho,\Gamma)$, is a convex function of
$\eta$ (positive second derivative), and decreasing for $\eta\leq1$
(negative first derivative). Thus, the infimum over $\eta\geq1$ is
achieved by the point where the derivative w.r.t. $\eta$ of the objective
function in \eqref{eq: Phi definition} vanishes. After some straightforward
algebra, we obtain that the optimal $\eta^{*}$ is the larger solution
of the quadratic equation
\begin{equation}
(\rho+1)^{2}\eta^{2}-\left[(\rho+1)(2\rho+1)+\Gamma\right]\eta+\rho^{2}+1=0\label{eq: optimal eta quadratic equation}
\end{equation}
given by \eqref{eq: eta star}.
\begin{rem}
The inequality $1-(1-\alpha)^{M}\leq(M\alpha)^{\rho}$ for $0\leq\rho\leq1$
can be replaced with \cite[Lemma 1]{Anelia_fingerprinting}
\begin{equation}
\frac{1}{2}\min\{1,M\alpha\}\leq1-(1-\alpha)^{M}\leq\min\{1,M\alpha\}\label{eq: clipped union bound}
\end{equation}
which states that the union bound, when clipped to $1$, is asymptotically
tight. Our analysis above can also be carried out using \eqref{eq: clipped union bound}
in \eqref{eq: error probability first step}, to obtain the exact
exponential behavior of the error probability. However, the resulting
expressions are more complicated, and we have not found any specific
cases for which the numerical value of the bound derived with \eqref{eq: clipped union bound}
is better than the bound derived above.
\end{rem}

\section{Proofs for Asymptotic SNR Analysis\label{sec:SNR-Asymptotics }}
\begin{IEEEproof}[Proof of Prop. \ref{prop: High SNR constant}]
First, we approximate Gallager's function in the regime $\Gamma\gg1$.
We have 
\begin{align}
\beta_{0} & =\frac{1}{2}\left(1+\frac{\Gamma}{1+\rho}\right)\left[1+\sqrt{1-\frac{4\Gamma\rho}{(1+\rho+\Gamma)^{2}}}\right]\\
 & =\frac{1}{2}\left(1+\frac{\Gamma}{1+\rho}\right)\left[1+1-\frac{2\Gamma\rho}{(1+\rho+\Gamma)^{2}}+\Theta\left(\frac{1}{\Gamma^{2}}\right)\right]\\
 & =\left(1+\frac{\Gamma}{1+\rho}\right)\left[1-\frac{\Gamma\rho}{(1+\rho+\Gamma)^{2}}+\Theta\left(\frac{1}{\Gamma^{2}}\right)\right]\\
 & =1+\frac{\Gamma}{1+\rho}-\frac{\rho}{1+\rho}\cdot\frac{\Gamma^{2}}{(1+\rho+\Gamma)^{2}}+\Theta\left(\frac{1}{\Gamma}\right)\\
 & =\frac{1+\Gamma}{1+\rho}+o(\Gamma).\label{eq: beta_0 high SNR}
\end{align}
for \eqref{eq: beta_0}, and
\begin{align}
E_{0}(\rho,\Gamma) & =(1-\beta_{0})(1+\rho)+\Gamma+\log\left(\beta_{0}-\frac{\Gamma}{1+\rho}\right)+\rho\log(\beta_{0})\\
 & =\rho+\log\left(\frac{1}{1+\rho}\right)+\rho\log\left(\frac{1+\Gamma}{1+\rho}\right)+o(\Gamma),\\
 & =\rho-(1+\rho)\log(1+\rho)+\rho\log(\Gamma)+o(\Gamma),\label{eq: E0 high SNR}
\end{align}
for \eqref{eq: E0}. 

For the channel coding converse bound of Proposition \ref{prop:Channel coding upper bound},
observing \eqref{eq: E0 high SNR}, it evident that the minimum in
\eqref{eq: Channel Coding Upper Bound} is attained by $E_{0}(\alpha,\Gamma)$
for high SNR, which leads directly to the first case in \eqref{eq: high SNR costant}.
The spherical cap bound of Theorem \ref{thm:Spherical cap bound}
at high SNR simply reads
\[
F_{\alpha}(\Gamma)\leq\alpha\log(\Gamma)+\alpha\log\left(\frac{\gamma_{\alpha}}{\alpha}\right)+\alpha.
\]

It remains to analyze the behavior of the spectrum replication bound
(Theorem \ref{thm: spectrum replication bound}) for high SNR ($\Gamma\to\infty$).
Approximating \eqref{eq: eta star}, we get 
\begin{align}
\eta^{*} & =\frac{\Gamma+\sqrt{\Gamma^{2}-4(\rho^{2}+1)(\rho+1)^{2}}}{2(\rho+1)^{2}}\\
 & =\frac{\Gamma+\Gamma\sqrt{1-\nicefrac{4(\rho^{2}+1)(\rho+1)^{2}}{\Gamma^{2}}}}{2(\rho+1)^{2}}\\
 & =\frac{\Gamma+\Gamma\left[1-\nicefrac{2(\rho^{2}+1)(\rho+1)^{2}}{\Gamma^{2}}+\Theta\left(\frac{1}{\Gamma^{4}}\right)\right]}{2(\rho+1)^{2}}\\
 & =\frac{\Gamma}{(\rho+1)^{2}}+\Theta\left(\frac{1}{\Gamma}\right).
\end{align}
Inserting back to \eqref{eq: Phi definition}, we get 
\[
\Phi(\rho,\Gamma)=\rho\left[\frac{\Gamma}{(\rho+1)^{2}}-1-\log\left(\frac{\Gamma}{(\rho+1)^{2}}\right)\right]+\frac{\Gamma}{(\rho+1)^{2}}+\Gamma+\log(\rho+1)-\frac{2\Gamma}{\rho+1}+o(\Gamma).
\]
Then, 
\begin{align}
\Lambda_{\alpha}(\Gamma) & =\sup_{0<\rho\leq1}\frac{\Phi(\rho,\Gamma)-\gamma_{\alpha}\Gamma}{\rho}\\
 & =\sup_{0<\rho\leq1}\frac{\Gamma}{(\rho+1)^{2}}-1-\log\left(\frac{\Gamma}{(\rho+1)^{2}}\right)+\frac{\Gamma}{\rho(\rho+1)^{2}}+\nonumber \\
 & \hphantom{=}\frac{\Gamma}{\rho}+\frac{\log(\rho+1)}{\rho}-\frac{2\Gamma}{\rho(\rho+1)}-\frac{\gamma_{\alpha}\Gamma}{\rho}+o(\Gamma)\\
 & =\sup_{0<\rho\leq1}\left[\frac{1}{(\rho+1)^{2}}+\frac{1}{\rho(\rho+1)^{2}}+\frac{(1-\gamma_{\alpha})}{\rho}-\frac{2}{\rho(\rho+1)}\right]\Gamma-\nonumber \\
 & \hphantom{=}\log\left[\frac{\Gamma}{(\rho+1)^{2}}\right]-1+\frac{\log(\rho+1)}{\rho}+o(\Gamma)\\
 & =\sup_{0<\rho\leq1}\left[\frac{\rho+1+(1-\gamma_{\alpha})(\rho+1)^{2}-2(\rho+1)}{\rho(\rho+1)^{2}}\right]\Gamma\nonumber \\
 & \hphantom{=}-\log\left[\frac{\Gamma}{(\rho+1)^{2}}\right]+\frac{\log(\rho+1)}{\rho}-1+o(\Gamma)\\
 & =\sup_{0<\rho\leq1}\left[\frac{-\rho-1+(1-\gamma_{\alpha})(\rho+1)^{2}}{\rho(\rho+1)^{2}}\right]\Gamma-\log\left[\frac{\Gamma}{(\rho+1)^{2}}\right]+\frac{\log(\rho+1)}{\rho}-1+o(\Gamma)\\
 & =\sup_{0<\rho\leq1}\left[\frac{(1-\gamma_{\alpha})\rho^{2}+(1-2\gamma_{\alpha})\rho-\gamma_{\alpha}}{\rho(\rho+1)^{2}}\right]\Gamma-\log\left[\frac{\Gamma}{(\rho+1)^{2}}\right]+\frac{\log(\rho+1)}{\rho}-1+o(\Gamma).
\end{align}
Clearly, for $\Gamma\to\infty$ the maximizer $\rho$ is chosen to
maximize the coefficient of the linear dependence on $\Gamma$. Differentiating
this coefficient w.r.t. $\rho$, we get
\[
\frac{(\gamma_{\alpha}-1)\rho^{3}+(3\gamma_{\alpha}-1)\rho^{2}+3\gamma_{\alpha}\rho+\gamma_{\alpha}}{\rho^{2}(1+\rho)^{3}},
\]
and when this derivative is strictly positive for all $\rho\in(0,1]$,
the supremum is attained for $\rho=1$. It can be verified (e.g.,
numerically) that this happens as long as $\gamma_{\alpha}\gtrsim0.0175$.
If we use the bound \eqref{eq: Burnashev bound} instead of the actual
value of $\gamma_{\alpha}$, then this results that $\rho=1$ is optimal
for all $\alpha\gtrsim0.0178$. In all these cases, we get
\[
\Lambda_{\alpha}(\Gamma)=\left[\frac{1}{2}-\gamma_{\alpha}\right]\Gamma-\log(\Gamma)+\log8-1+o(\Gamma),
\]
and inserting back to \eqref{eq: Superposition Bound} we get the
bound 
\begin{equation}
F_{\alpha}^ {}(\Gamma)\leq(1+\alpha)\left[\gamma_{\alpha}-\frac{\alpha}{2(1+\alpha)}\right]\Gamma+\alpha\log(\Gamma)-\alpha\log\left(\frac{8}{e}\right)+o(\Gamma).\label{eq: high SNR spherical cap}
\end{equation}
Further, for $\alpha\geq2$, using the expression in \eqref{eq: Burnashev bound}
we have $\gamma_{\alpha}=\frac{\alpha}{2(1+\alpha)}$ which implies
\[
F_{\alpha}(\Gamma)\leq\alpha\log(\Gamma)-\alpha\log\left(\frac{8}{e}\right)+o(\Gamma).
\]
\end{IEEEproof}
\begin{rem}
\label{rem:Conjecture based on asymptotic analysis}If one can prove
that the reverse inequality in \eqref{eq: high SNR spherical cap}
holds, i.e., 
\[
F_{\alpha}(\Gamma)\geq(1+\alpha)\left[\gamma_{\alpha}-\frac{\alpha}{2(1+\alpha)}\right]\Gamma+\alpha\log(\Gamma)-\alpha\log\left(\frac{8}{e}\right)+o(\Gamma),
\]
then this could lead to stronger results for the\emph{ unlimited-bandwidth
case}, showing that $\gamma_{\alpha}=\frac{\alpha}{2(1+\alpha)}$
for all $\alpha$ (rather than $\alpha\geq2$, as was previously known),
along with a simpler proof than \cite{Burnashev85} (albeit somewhat
indirect). Indeed, if $\gamma_{\alpha}>\frac{\alpha}{2(1+\alpha)}$
then $F_{\alpha}^ {}(\Gamma)$ would increase linearly with $\Gamma$,
which is clearly unacceptable. To obtain a contradiction, one can
derive and channel encoder and decoder by a proper quantization of
the optimal modulator and estimator, and show that the communication
rate increases linearly with $\Gamma$ with a negligible error probability.
This evidently contradicts the logarithmic behavior of the capacity
in $\Gamma$. The main gap in such a proof method, however, is to
show a reverse inequality in \eqref{eq: min of two exponent  upper bound}.
In turn, this corresponds to the hypothesis that the unlimited-bandwidth
system constructed in the proof of the spectrum replication bound
is asymptotically optimal. Even more specifically, it seems difficult
to argue why the restriction of the estimator to a two steps procedure
is asymptotically optimal.\end{rem}
\begin{IEEEproof}[Proof of Prop. \ref{prop:High SNR behviour channel coding lower bound}]
Using the approximations for $\beta_{0}$ and $E_{0}(\rho,\Gamma)$
in \eqref{eq: beta_0 high SNR} and \eqref{eq: E0 high SNR}, the
first term in \eqref{eq: Channel Coding Lower Bound} is approximated
as
\[
\sup_{0\leq\rho\leq1}\frac{\alpha E_{0}(\rho,\Gamma)}{\rho+\alpha}=\sup_{0\leq\rho\leq1}\frac{\alpha\left[\rho-(1+\rho)\log(1+\rho)+\rho\log\Gamma+o(\Gamma)\right]}{\rho+\alpha}.
\]
At high SNR, the optimal choice for $\rho$ is the one maximizing
the coefficient of $\log(\Gamma)$, i.e. $\frac{\alpha\rho}{\rho+\alpha}$
which is $\rho=1$. This results the lower bound
\begin{align}
 & \hphantom{=}\frac{\alpha}{1+\alpha}\left[1-2\log2+\log\Gamma+o(\Gamma)\right]\nonumber \\
 & =\frac{\alpha}{1+\alpha}\cdot\left[1+o(\Gamma)\right]\cdot\log\Gamma.\label{eq:  high SNR E0 term asymptotics}
\end{align}
Now, let us inspect the second term in \eqref{eq: Channel Coding Lower Bound},
i.e., 
\begin{equation}
\sup_{\rho\geq1}\frac{\alpha E_{\st[x]}(\rho,\Gamma)}{\rho+\alpha}.\label{eq: channel coding lower bound- expurgated term in high SNR proof}
\end{equation}
Let $\rho_{\st[x]}^{*}(\Gamma)$ be the maximizing value of $\rho$.
Consider the hypothesis that $\rho_{\st[x]}^{*}(\Gamma)$ increases
linearly with $\Gamma$. Then, the denominator of \eqref{eq: channel coding lower bound- expurgated term in high SNR proof}
increases linearly with $\Gamma$. It can easily be seen that the
nominator cannot increase faster then linear, and so the value of
\eqref{eq: channel coding lower bound- expurgated term in high SNR proof}
is bounded as $\Gamma\to\infty$. Such a behavior is of course unreasonable,
and as will shall see, better value for \eqref{eq: channel coding lower bound- expurgated term in high SNR proof}
can be attained. Next, consider the hypothesis that $\rho_{\st[x]}^{*}(\Gamma)$
increases sub-linearly with $\Gamma$, which implies that $\frac{\Gamma}{\rho_{\st[x]}^{*}(\Gamma)}\to\infty$
as $\Gamma\to\infty$. In this event, 
\begin{align}
\beta_{\st[x]} & =\frac{1}{2}+\frac{\Gamma}{4\rho_{\st[x]}^{*}(\Gamma)}+\frac{1}{2}\sqrt{1+\frac{\Gamma^{2}}{4\rho_{\st[x]}^{*}(\Gamma)^{2}}}\\
 & =\frac{1}{2}+\frac{\Gamma}{2\rho_{\st[x]}^{*}(\Gamma)}+o(\Gamma),
\end{align}
and then the objective in \eqref{eq: channel coding lower bound- expurgated term in high SNR proof}
is approximated as
\begin{align}
 & \hphantom{=}\frac{\alpha}{\rho_{\st[x]}^{*}(\Gamma)+\alpha}\left[2(1-\beta_{\st[x]})\rho+\Gamma+\rho\log\left[\beta_{\st[x]}\left(\beta_{\st[x]}-\frac{\Gamma}{2\rho}\right)\right]\right]\nonumber \\
 & =\frac{\alpha}{\rho_{\st[x]}^{*}(\Gamma)+\alpha}\left[\rho_{\st[x]}^{*}(\Gamma)+\rho_{\st[x]}^{*}(\Gamma)\cdot\log\left(\frac{1}{4}\right)+\rho_{\st[x]}^{*}(\Gamma)\cdot\log\left[1+\frac{\Gamma}{\rho_{\st[x]}^{*}(\Gamma)}\right]\right]+o(\Gamma)\\
 & =\frac{\alpha\cdot\rho_{\st[x]}^{*}(\Gamma)}{\rho_{\st[x]}^{*}(\Gamma)+\alpha}\cdot\log\left[\frac{e\cdot\Gamma}{4\rho_{\st[x]}^{*}(\Gamma)}\right]+o(\Gamma).
\end{align}
Now, if $\rho_{\st[x]}^{*}(\Gamma)=\Gamma^{\nu(\Gamma)}$ for some
function $\nu(\Gamma)$ such that $\nu(\Gamma)\to0$ yet $\rho_{\st[x]}^{*}(\Gamma)\to\infty$
as $\Gamma\to\infty$ (e.g. $\nu(\Gamma)=\log(\Gamma)$) then last
expression is asymptotically given by 
\[
\alpha\cdot\left[1+o(\Gamma)\right]\cdot\log\Gamma.
\]
Comparing the last expression with \eqref{eq:  high SNR E0 term asymptotics}
it is apparent that as $\Gamma\to\infty$ the expurgated term dominates
the maximization of \eqref{eq: Channel Coding Lower Bound}, and the
bound scales as claimed.
\end{IEEEproof}

\begin{IEEEproof}[Proof of Prop. \ref{prop:Low SNR behviour channel coding lower bound}]
As we are interested in $\Gamma\to0$ we may clearly assume that
$\Gamma<1$ and so, e.g., $\Theta(\Gamma^{2})+\Theta(\Gamma)=\Theta(\Gamma)$.
First, we approximate Gallager's function. We have 
\begin{align}
\beta_{0} & =\frac{1}{2}\left(1+\frac{\Gamma}{1+\rho}\right)\left[1+\sqrt{1-\frac{4\Gamma\rho}{(1+\rho+\Gamma)^{2}}}\right]\\
 & =\frac{1}{2}\left(1+\frac{\Gamma}{1+\rho}\right)\left[2-\frac{2\Gamma\rho}{(1+\rho+\Gamma)^{2}}+\Theta(\Gamma^{2})\right]\\
 & =1+\frac{\Gamma}{1+\rho}-\frac{\Gamma\rho}{(1+\rho+\Gamma)^{2}}+\Theta(\Gamma^{2})\\
 & =1+\frac{\Gamma}{1+\rho}-\frac{\Gamma\rho}{(1+\rho)^{2}}+\Theta(\Gamma^{2})\\
 & =1+\frac{1}{(1+\rho)^{2}}\Gamma+\Theta(\Gamma^{2})\label{eq: beta_0 low SNR}
\end{align}
for \eqref{eq: beta_0}, and
\begin{align}
E_{0}(\rho,\Gamma) & =(1-\beta_{0})(1+\rho)+\Gamma+\log\left(\beta_{0}-\frac{\Gamma}{1+\rho}\right)+\rho\log\beta_{0}\\
 & =-\frac{1}{(1+\rho)}\Gamma+\Gamma+\log\left[1+\frac{1}{(1+\rho)^{2}}\Gamma-\frac{\Gamma}{1+\rho}\right]+\rho\log\left[1+\frac{1}{(1+\rho)^{2}}\Gamma\right]+\Theta(\Gamma^{2})\\
 & =\frac{\rho}{(1+\rho)}\Gamma+\log\left[1-\frac{\rho}{(1+\rho)^{2}}\Gamma\right]+\rho\log\left[1+\frac{1}{(1+\rho)^{2}}\Gamma\right]+\Theta(\Gamma^{2})\\
 & =\frac{\rho}{(1+\rho)}\Gamma-\frac{\rho}{(1+\rho)^{2}}\Gamma+\frac{\rho}{(1+\rho)^{2}}\Gamma+\Theta(\Gamma^{2})\\
 & =\frac{\rho}{(1+\rho)}\Gamma+\Theta(\Gamma^{2}),\label{eq: E0 Low SNR}
\end{align}
for \eqref{eq: E0}. Thus, the first term in \eqref{eq: Channel Coding Lower Bound}
is approximated as
\[
\sup_{0\leq\rho\leq1}\frac{\alpha E_{0}(\rho,\Gamma)}{\rho+\alpha}=\sup_{0\leq\rho\leq1}\frac{\alpha\rho\Gamma}{(1+\rho)(\rho+\alpha)}+\Theta(\Gamma^{2}),
\]
and clearly, the maximizer $\rho$ is the one maximizing $\frac{\alpha\rho}{(1+\rho)(\rho+\alpha)}$,
i.e. $\rho=1$. Hence, the first term is
\[
\frac{\alpha\Gamma}{2(1+\alpha)}+\Theta(\Gamma^{2}).
\]
Analyzing the second term in \eqref{eq: Channel Coding Lower Bound}
only leads to a worse behavior and thus may be disregarded.
\end{IEEEproof}

\bibliographystyle{plain}
\bibliography{Modulation_Estimation}

\end{document}